\documentclass[11pt,a4paper]{article}
\usepackage{amsfonts}
\usepackage{amssymb}
\usepackage{jheppub}
\usepackage{ulem}
\usepackage{extarrows}
\usepackage{multirow}
\usepackage{float}

\usepackage{inputenc}
\usepackage{xspace}
\usepackage{url}

\def\bea{\begin{eqnarray}}
\def\eea{\end{eqnarray}}
\newcommand{\be}{\begin{equation}}
\newcommand{\ee}{\end{equation}}

\def\braket#1{\mathinner{\langle{#1}\rangle}}
\newcommand{\sbraket}[1]{\lbrack #1\rbrack}

\newcommand{\boxit}[1]{%
  \[\fbox{%
      \addtolength{\linewidth}{-2\fboxsep}%
      \addtolength{\linewidth}{-2\fboxrule}%
      \begin{minipage}{\linewidth}%
      #1%
      \end{minipage}%
    } \nonumber \]%
}

\title{On-shell recursion relations for generic integrands}
\author[a]{Rutger H. Boels}
\author[a,b]{and Hui Luo}
\affiliation[a]{II. Institut f\"ur Theoretische Physik, Universit\"at Hamburg\\ Luruper Chaussee 149, D- 22761 Hamburg, Germany }
\affiliation[b]{Kavli Institute for Theoretical Physics China\\ CAS, Beijing 100190, China}

\keywords{Amplitudes}

\abstract{The quantum effects encapsulated in loop corrections are crucial in quantum field theory for a wide variety of formal and phenomenological applications. In this article we propose and check a definition of the so-called single cut contributions needed to complete on-shell recursion relations for the integrand of scattering amplitudes in generic power-counting renormalisable theories at conjecturally any loop order. Our proposal meshes well with standard dimensional regularisation and applies in particular directly to much of the standard model of particle physics. Apart from a diagrammatic construction, at one loop order we provide a direct cross-check for box coefficients. Interestingly, at one loop our proposal can be related to a specific subset of all double unitarity cuts as well as tree-level poles by iterated recursion. We focus in particular on demonstrating the method in rational-term one-loop examples in pure Yang-Mills theory. For the finite amplitudes we present all-order arguments. First steps toward gravity integrands are taken.}

\begin{document}
\maketitle

\section{Introduction}
Perturbative computation is the backbone of particle physics as a predictive science. Typically, observables are computed as a perturbative series in a (set of) coupling constant(s). This translates into the loop expansion of a quantum field theory. Computing these quantum corrections to collider relevant observables is of near-universal interest. Moreover, quantum corrections are important as well beyond direct phenomenology applications for insight into the structure of quantum field theory. This is also important for more formal applications of QFT. 

The text-book road to perturbative computation is through Feynman graphs. Its advantages are clear: perturbative unitarity and locality are manifest, as well as Lorentz symmetry and gauge symmetry in dimensional regularisation. However, apart from computational complexity this particular representation may obscure symmetries of the answer which are not manifest at the Lagrangian level. In recent years several such examples have been found by focusing on the full results for scattering amplitudes, instead of their Feynman graph representation, see e.g. \cite{Elvang:2013cua} \cite{Henn:2014yza}. This has in turn lead to a renewed focus on methods to compute perturbative quantities beyond Feynman graphs. 

An important example of such methods are the on-shell recursion relations for tree level amplitudes of Britto, Cachazo, Feng and Witten \cite{Britto:2004ap, Britto:2005fq}. These relations allow one to compute scattering amplitudes from a subset of the residues at kinematic singularities. Since these singularities are governed by unitarity at tree level, this gives a canonical construction of scattering amplitudes. On-shell relations for tree level amplitudes exist in basically any power counting renormalisable Yang-Mills theory as well as many Einstein-gravity theories in four or more dimensions, see \cite{ArkaniHamed:2008yf}. BCFW on-shell recursion relations have been extended in many directions, see for instance  \cite{Feng:2011np} \cite{Elvang:2013cua} \cite{Henn:2014yza} and references therein. 

Compared to understanding at tree level, developments at loop level have been much less advanced. Fuelled by the tree level progress many loop level techniques have been explored, all of which recycle tree amplitudes into loop amplitudes. The most powerful and influential of these to date involve a breakdown of the one loop level scattering amplitudes into a known integral basis, with coefficients left to be determined. This has lead in recent years to next-to-leading order (NLO) as the regular benchmark of collider process computations for cross-sections. The approach through an integral basis originates in older work, e.g. in \cite{Bern:1994cg}, and has been extended beyond the one loop order with a degree of success, especially for the planar sector of $\mathcal{N}=4$ supersymmetric gauge theories \cite{Bern:2005iz}. Typically, one constructs a large enough Ansatz which is then whittled down to the physical answer by applying unitarity cut constraints. Although very powerful, this approach is not completely algorithmic as it requires for instance knowledge of the integral basis which is not always available. Also, this approach does not give much structural insight into quantum corrections in the most general, phenomenologically relevant models beyond one loop. In this letter we explore the extension of on-shell recursion relations into this direction for the integrand of a class of theories. 

On-shell recursion relations for the integrand of planar $\mathcal{N}=4$ super Yang-Mills theories in strictly four dimensions were proposed in  \cite{ArkaniHamed:2010kv}. Although it was simultaneously proven \cite{Boels:2010nw} that the integrand is determined by certain kinematic singularities for power-counting renormalisable gauge theories, these singularities are not directly determined by unitarity or by symmetry. The latter is the case for unregulated $\mathcal{N}=4$. The obstacle to extending to generic gauge theories is the definition of so-called single-cut singularities. An added complication is the fact that the integrand is only defined up to terms which integrate to zero. While for planar $\mathcal{N}=4$ in strictly four dimensions the answer is fixed by symmetries, even there the integrand cannot be integrated without further regularisation due to IR divergences. To overcome these obstacles in this article we propose and motivate a generic construction of the single-cut singularities, applicable to quite generic $D$-dimensional QFTs. The proposal is verified in a number of example computations against known results in the literature.

The reader should not get the impression from the above that on-shell recursion is the only game in town for recycling tree amplitudes into loop amplitudes. The oldest known example is probably Feynman's tree theorem \cite{Feynman:1963ax}, which expresses a loop amplitude as a sum over an at least singly cut (!) set of graphs. Interestingly, this method has been applied to full string theory \cite{Brink:1973gi} in the early days of the subject, see \cite{Brandhuber:2005kd} for a modern application. Feynman's tree theorem has been generalised recently to include only single-cut terms, see e.g. \cite{Catani:2008xa}, and in other directions \cite{CaronHuot:2010zt}. A computation using single cuts and the massive CSW formalism has appeared in \cite{NigelGlover:2008ur}. Recently an attempt was made to combine BCFW and Feynman's tree theorem \cite{Maniatis:2016gui} \cite{Maniatis:2016nmc}, with some result for a one loop, two legs on-shell form factor in QED. Earlier attempts include a master thesis written under direction of one of the present authors, \cite{hanssenthesis}. String theory inspiration takes another more recent appearance in the so-called CHY representation of tree amplitudes \cite{Cachazo:2013hca}, closely related to the existence of novel twistor string theories \cite{Mason:2013sva}. This approach has been extended to loop level, conjecturally to all orders, by employing twistor string methods at loop level \cite{Geyer:2015jch},or by more direct methods \cite{Cachazo:2015aol},\cite{Feng:2016nrf}. Single cut terms in the CHY formalism were studied in \cite{He:2015yua}. Further exciting developments relating trees and loops are the so-called Q-cut representation \cite{Baadsgaard:2015twa} \cite{Huang:2015cwh}. Using so-called on-shell diagrams which encode four dimensional amplitudes a partially successful attempt at loop level computation in less-than-$\mathcal{N}=4$ supersymmetric gauge theories was made in \cite{Benincasa:2015zna}, which covers certain cut-constructible terms. 

While the mentioned works share similarities and certainly philosophy with the drive of the current article, they are technically distinct. Our approach is less high-tech, but has the distinct advantage is that it quite directly applies to a large part of the standard model of particle physics, or more generally to the class of power-counting renormalisable gauge theories, as well as gravity to an extent. Moreover, standard dimensional regularisation can be used easily, without the need for regularisation on a case-by-case basis. In view of the usual calculational chain toward LHC-ready cross-sections, this is strongly desired. Our approach will be demonstrated in section \ref{sec:intro}. The main result is contained in equation \eqref{eq:masterformulasinglecutsYM} . Apart from diagrammatics, a quite general argument will be given that our construction yields the right integrand, certainly to the one loop order. This claim will be verified in explicit examples in section three. Our explicit results generically share a nasty feature with all the previously mentioned recent developments: the generated integrands have spurious singularities which generally only vanish after integration. Also in line with most recent advances, we will not move beyond the one loop order in this article. It is worth emphasising though that the current frontier of experimentally relevant amplitudes is at two loops, and therefore not very far away. In section four we show how our results apply to the simplest non-supersymmetric and maximally supersymmetric gravitational examples. We end with a discussion and conclusions. Several appendices deal with technical details of the computations.

\textit{Note added in proof}
While this article was being readied for publication, the interesting article \cite{scoopdiedoop} appeared on the arXiv which has some overlap with the drive an results of the present article.

\section{On-shell recursion at loop level: generics}\label{sec:intro}

The standard BCFW shift for two legs reads
\begin{equation}\label{eq:bcfwshift}
p_1 \rightarrow p_1 + q \, z \qquad p_n \rightarrow p_n - q \, z
\end{equation}
where $p_1 \cdot q = p_n \cdot q = q \cdot q = 0$. Under the latter condition, the masses of the shifted legs remain invariant. Moreover, momentum conservation is guaranteed. As shown in \cite{Boels:2010nw}, the planar integrand of Yang-Mills gauge theories, minimally coupled to matter behaves under the shift of two gluons as,
\begin{equation}\label{eq:bcfwshiftint}
\lim_{z\rightarrow \infty} I(z) \propto \hat{\xi_1}^{\mu} \hat{\xi_2}^{\nu} \left (z \, \eta^{\mu\nu} + B_{\mu\nu} + \mathcal{O} \left(\frac{1}{z} \right) \right)
\end{equation}
for some anti-symmetric matrix $B$. This handy form of the large $z$ behavior was first found for tree amplitudes in \cite{ArkaniHamed:2008yf}, using background field methods. Introducing a contour integral around $z=0$ and pushing this to infinity allows one to derive
\begin{equation}
I(0) = \oint_{z=0} \frac{I(z)}{z} = - \sum_{i} \textrm{Res}_i I(z=\textrm{finite}) - \textrm{Res} (z=\infty)
\end{equation}
For each helicity choice of the external shifted gluons, a shift can be found such that the residue at infinity drops out and a relation is obtained which expresses the integrand as a sum over a subset of its singularities. For generic momenta, a simple inspection of the Feynman graphs or a generic argument through unitarity shows that the most complicated singularity to arise for generic external momenta is a simple pole. If the residues at these poles are known, the integrand can be computed. 

At tree level, all singularities are physical and their residues are governed by unitarity to be the product of tree level amplitudes. At loop level, the loop integration invalidates the same conclusion for the integrand, at least naively: cutting a single loop-momentum dependent leg is not governed by unitarity. At the loop level, cutting at least two loop legs is governed through branch cut singularities by Cutkosky rules \cite{Cutkosky:1960sp}: this yields the Lorentz-invariant phase space integral over the product of two lower-loop amplitudes, with the cut-legs on-shell. In recent years, generalised unitarity techniques have been applied to cutting more than two legs. By inspection of Feynman graphs, these generalised unitarity cuts correspond to products of lower loop amplitudes. Important for this paper though is the observation that the problem of writing down effective recursion relations at the loop level reduces to finding a definition of the so-called `single cut' term. As noted in the introduction, the concept of singly cut term has appeared before in several places. 

The integrand is inherently an ill-defined object: only the integrated expression has a physical meaning. On a practical level, this problem appears as the possibility to shift the loop momentum without changing the value of the integral. As a case in point, consider the massless tadpole integral which vanishes in dimensional regularisation 
\begin{equation}
\int d^D l \frac{1}{l^2} = 0 
\end{equation}
Basically, as long as the integral is well-defined, it can only vanish: it is a scale-less integral which has a non-zero mass-dimension. Consider this integrand under BCFW on-shell recursion: a pole at infinity arises and on-shell recursion does not even work directly, although the end-result is trivial. The same fate can befall finite shift residues. Consider for this massless tadpole integral, shifted by a momentum $p_1$
\begin{equation}
\int d^D l \frac{1}{(l+p_1)^2} = 0 
\end{equation}
Under the BCFW shift in equation \eqref{eq:bcfwshift}, this integrand has no residue at infinity, but instead has acquired a finite $z$ residue, which is non-trivial. It of course still vanishes after integration. 

The issue of loop integration ambiguity for the integrand affects the definition of the BCFW on-shell recursion relations for the integrand as well: per Feynman graph a specific momentum routing needs to be specified. This problem was solved for gluon amplitudes in \cite{Boels:2010nw} by picking $z$-dependence of the loop momenta propagators to follow the path of shortest length through individual Feynman graphs in color-ordered Yang-Mills amplitudes in q-lightcone gauge. If two color-adjacent particles are shifted, this implies to the one-loop order that the singularity can be taken to be contained in a single propagator:
\begin{equation}\label{eq:cutloop}
(l+\hat{p}_1)^2 = 0 
\end{equation}
at least up to a $z$-independent shift. For higher loops, more complicated possibilities arise.

This leads to the following setup for loop level on-shell recursion relations:
\begin{equation}
I(0) = \oint_{z=0} \frac{I(z)}{z} = - \sum_{i} \textrm{Res}_i I(\textrm{tree-like}) - I(\textrm{loop-like})
\end{equation}
where the tree-like singularity residues are the ones where a tree-level propagator goes on-shell. This is still governed by unitarity. At one loop order, the other 'single-cut' term reads
\begin{equation}
I(\textrm{loop-like}) = \frac{1}{(l+p_1)^2} \tilde{I}_{\textrm{ll}} (l_1, l_2, \hat{p_1} p_2 \ldots p_{n-1} \hat{p_n} )
\end{equation}
where $z$ is the solution to equation \eqref{eq:cutloop} and 
\begin{equation}
l_1 = -l_2 = l+ \hat{p_1}
\end{equation}
The problem of finding the single cut integrand is now specialised to the problem of defining $\tilde{I}_{\textrm{ll}}$ with this specific loop momenta for the loop legs. We want to construct it out of lower order scattering amplitudes with two additional particles. 

\begin{figure}[!htb]
\centering
\includegraphics[scale=0.5]{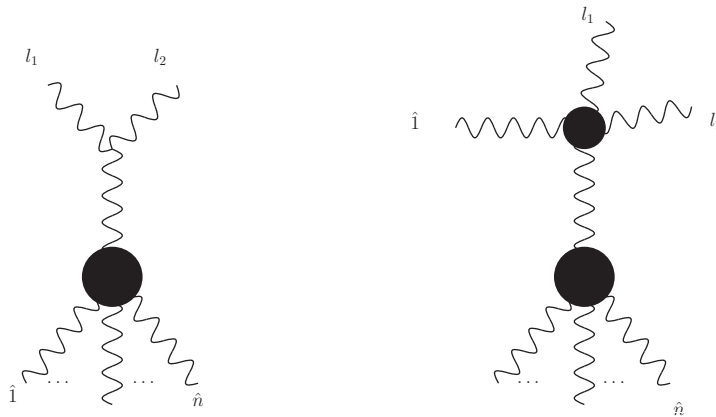}
\caption{Difference in Feynman graphs between single cuts and lower order scattering amplitudes with additional legs. There is an additional graph with a bubble on the other shifted leg.}
\label{fig:feyngraphextra}
\end{figure}

Note that the singly cut integrand has at least one proper definition: this is inherited from the q-lightcone gauge Feynman graphs used to derive \eqref{eq:bcfwshiftint}! It is clear without much computation that many of the Feynman graphs correspond to those for scattering amplitudes with two additional particles. From this point we will specialise to color-ordered loop amplitudes, with an adjacent BCFW shift. There are two obvious differences between the color-ordered Feynman graphs of the scattering amplitudes with additional particles and the singly cut integrand: the classes of graphs denoted in the figure \ref{fig:feyngraphextra} are present in the scattering amplitude, but not in the singly-cut integrand. The left graph looks like a tadpole contribution, while the right graph resembles an external leg correction (there is a similar graph on the other shifted leg). Both these classes of Feynman graphs would not be included in the Feynman graph representation of the integrand as they correspond to vanishing integrals.

It pays to discuss quantum numbers explicitly at this point. Unobserved quantum numbers after all need to be summed over. The helicity quantum numbers of the gluons for instance need to be summed over to get the right dependence on the little group indices of the external gluons. Furthermore, we will mostly consider color-ordered integrands for planar amplitudes in Yang-Mills theories. The would-be integrand with two additional gluons would depend on more color quantum numbers than the sought-for integrand. The only rational, gauge invariant and physical choice is to sum over these. For color-ordered amplitudes, this means that there are two color-orderings of the loop legs to sum over: both contribute equally to the color-ordered integrand. 
\begin{equation}\label{eq:guess1}
\tilde{I}_{\textrm{ll}}  \stackrel{?}{=} A_{n+2} (\stackrel{\longleftrightarrow}{l_1, l_2}, \hat{p_1} p_2 \ldots p_{n-1} \hat{p_n} )
\end{equation}
This immediately solves a problem, emerging from the left diagram in Fig. \ref{fig:feyngraphextra}: color-ordered amplitudes are singular in the limit two adjacent legs have collinear momenta, $l_1 \propto l_2$. The residue of the singularity is proportional to the color-ordered three point amplitude. There are two functional forms for this amplitude possible: either helicity amplitudes $++-$ or $--+$, as generated in Yang-Mills theory, or the helicity equal amplitudes, which are generated by an $F^3$ interaction. For the latter, a dimensionful coupling constant is needed. Since this is absent in Yang-Mills, the latter case cannot arise. The MHV amplitude and it conjugate however are anti-symmetric under exchange of any two particles. Since gluons are boson, this means that one needs an antisymmetric color-factor, which is then uniquely determined to be the usual structure constant. Hence there are no singularities in the collinear limit to all loop orders, as long as the two color orders are summed over. At higher loop orders than one, this potentially only holds after integration.

The absence of collinear singularities does not make equation \eqref{eq:guess1} completely well-defined yet. Although the right hand side of equation \eqref{eq:guess1} is finite in the collinear limit, it's still divergent in the limit $l_1 \rightarrow -l_2$. The culprit is a soft divergence whenever the loop momentum legs connect directly to a line from an outside leg. In the case at hand, this occurs for instance for the two shifted legs. More generally, for shifts of two color-adjacent legs on a planar integrand one obtains a singularity for the two shifted legs, whose divergent propagator is of the form
\begin{equation}
\propto \frac{1}{(l_1 + l_2 + p_i)^2}
\end{equation}
These divergences very much look like external leg corrections, which, if they were exactly this, would vanish for on-shell external fields as the integrals are of the single-scale type. We will argue here these divergent contributions to the integrand always correspond to integrals which vanish after integration. 

To see the argument, consider a scattering amplitude multiplied by a loop momentum dependent factor,
\begin{equation}
\sim \frac{1}{(L+p_1)^2} \xi^1_{\mu} J^{\mu}\left(p_1 | p_2 \ldots, p_n \right)
\end{equation}
Here the amplitude has been written in terms of a current $J^{\mu}$, whose usually off-shell leg has been taken on-shell. This integrand vanishes after integration. Next, consider the BCFW shift \eqref{eq:bcfwshift} and write the recursion relation. For every propagator that goes on-shell inside the current, one obtains an integrand which manifestly integrates to zero. Then, the contribution to the integrand obtained by the loop momentum-containing propagator also integrates to zero. One can repeat this argument for
\begin{equation}\label{eq:oneexpansion}
\sim \frac{1}{(L+p_1)^2} \xi^1_{\mu} \left(L_{\nu} + p_{1,\nu} \right) \left(\frac{\partial}{\partial p_{\nu}}  J^{\mu}\left(p | p_2 \ldots, p_n \right) \right)_{p=p_1}
\end{equation}
In this case one could pick up a residue at infinity. This will however also integrate to zero: the loop momenta only appear in the prefactor in this case, which makes the $z=\infty$ contribution vanish.

To investigate the soft and collinear limits more closely, one can consider the following double BCFW shift,
\begin{align}
p_1 & = p_1 + z_1 q_1\\
l_1 & = l - z_1 q_1 \\
l_2 & = - l + z_2 q_2 \\
p_n & = p_n - z_2 q_2 
\end{align}
where $p_1 \cdot q_1 = l \cdot q_1 = q_1 \cdot q_1 = 0$ and  $p_n \cdot q_2 = l \cdot q_2 = q_2 \cdot q_2 = 0$, but also $q_1 \cdot q_2 \neq 0$. This shift has been constructed to obey momentum conservation.The system of constraints has two solutions. In the absence of collinear singularities, as is the case here, the limits $z_1\rightarrow 0$ and $z_2\rightarrow 0$ correspond to the soft singularities on the shifted legs. After a double Taylor expansion, a finite remainder is obtained beyond the singular term. The terms which appear have exactly the form of equation \eqref{eq:oneexpansion}.

The upshot is that one can safely ignore the soft divergences: they will all involve integrals which vanish after integration. On an operational level it means that throwing away terms which involve soft-divergent propagators will not change the value of the integrated expression. This means we could define the soft limit as for instance a contour integral, taking $w = z_1 = z_2$,
\begin{equation}\label{softlimit}
I = \int_{w=0} dw \frac{ A(X)}{w}
\end{equation}
which, up to terms which integrate to zero, gives the right integrand. Importantly, this quantity is \emph{gauge invariant}. Collecting results, our proposal for the single cut contribution to planar BCFW on-shell recursion relation is 
\boxit{
\begin{equation}\label{eq:masterformulasinglecutsYM}
I(l+ \hat{p_1}, X) = \oint_{w=0} dw \frac{1}{w} \sum_{\textrm{polarisations}} \lim_{l_1 \rightarrow (w-1)l_2} \sum_{color} A(l_1,l_2,X)
\end{equation}
}
with the added insight that one can ignore external bubble type divergences. We will verify below in examples this prescription and the added insight indeed yields the right answer. 

There is a simple extension to non-planar integrands which follows from the color considerations above. As a simple cross-check, if the above reasoning is applied to the subleading color amplitudes at one loop, the result obtained is directly consistent with the known KK-like relation for the one-loop integrals \cite{Bern:1994zx}. Note that the soft divergence in that case arise from the loop legs attaching to either side of a shifted leg. In general it would be interesting to see what relations for the integrand are consistent with the above definition, such as the ones first explored in \cite{Boels:2011mn}, see also \cite{Chester:2016ojq} \cite{Tourkine:2016bak}.

\subsection{Double recursion}
Using gauge invariance, it is easy to argue that the single-cut contribution as defined in equation \eqref{eq:masterformulasinglecutsYM} itself obeys BCFW on-shell recursion under a shift of any pair of legs. For a one-loop original integrand, after the first recursion step a tree-level object appears inside the single cut contribution, with additional tree-like factorisation contributions. Shifting two legs on the single cut contribution again will now only involve lower-order tree level objects. Some of these are tree-like factorisations again, while some involve cutting an additional loop leg. All tree-like factorisations involve lower loop integrands, while the doubly cut loop integrals can now be fixed by generalised unitarity as the product of two tree amplitudes in a certain momentum setup, up to terms which integrate to zero. In other words, if the single cut contribution indeed is given as in equation \eqref{eq:masterformulasinglecutsYM}, then the integrand obeys double recursion relations. If also the known physical answer obeys double recursion, then the validity of  equation \eqref{eq:masterformulasinglecutsYM} follows from unitarity. Double recursion was mentioned briefly in \cite{Boels:2010nw} for a particular shift of momenta. The computation below corresponds more closely to the case studied in \cite{gangandme}, which also featured a five point example in $\mathcal{N}=4$ SYM. 

For definiteness, let the first shift involve legs $1$ and $n$ and the second shift $2$ and $3$. The doubly cut terms involve among others the products of color-ordered tree amplitudes
\begin{equation}
A(\hat{1}, \tilde{2}, l_2, l_1 ) \, A(l_2, \tilde{3}, X, \hat{n}, l_1) \,\, . 
\end{equation}
Here $X$ stands for all non-shifted legs of the scattering amplitude. Taking different legs for the second shift will result in other, equivalent integral formulae. For the special case of four point scattering amplitudes, only the double cut contribution survives as the three point integrands will integrate to zero (this will be checked explicitly below).

\subsection{Cross-check for box integrals and their coefficients}
Above it was argued that the single cut contributions to the integrand obey on-shell recursion. If they do, unitarity will then basically fix the structure of all the residues to agree with the above proposal for the single cut in terms of lower-loop objects. Note that this will in general only hold up to terms which vanish after integration, such as the external bubble divergences. Hence the question is: is there a form of the integrand in which it is obvious the integrand's single cuts obey on-shell recursion? This will be checked to the one loop order for (massless and massive) box coefficients.

\subsubsection*{Kinematics of cut loop graphs}
The kinematics of cut loop graphs plays a central role in the analysis of the coefficients of box, bubble and triangle graphs. Here we follow basically \cite{Badger:2008cm}. 
\begin{figure}[t!]
\centering
\includegraphics[scale=0.8]{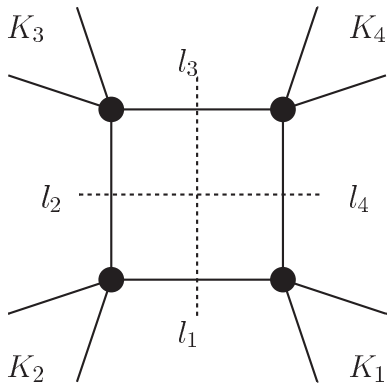}
\caption{\label{fig:quadruplecut} Momentum assignment of a massive box coefficient in a certain
channel. The $i$-th corner contain a number of gluons with total momentum $K_i$. Figure taken from \cite{Boels:2011mn}. }
\end{figure}

Given two massive momenta $K_1$ and $K_2$, consider the unique two massless momenta $K^{\flat}_1$ and $K^{\flat}_2$ such that
\begin{equation}
K_{1,\mu} = K^{\flat}_{1, \mu} + \frac{K_1^2}{\gamma_{12}}  K^{\flat}_{2, \mu}   \qquad K_{2,\mu} =  K^{\flat}_{2, \mu} + \frac{K_2^2}{\gamma_{12}}  K^{\flat}_{1, \mu} 
\end{equation}
with $\gamma_{12} = 2 K^{\flat}_1 \cdot K^{\flat}_2$. Their existence can be shown easily in the center of mass frame.  
Then, write a loop momentum as
\begin{equation}\label{l-decomposition}
l_{\mu} =  a K^{\flat}_{1,\mu} + b K^{\flat}_{2,\mu} + c n_{\mu} + d \overline{n}_{\mu}
\end{equation}
where $n, \overline{n}$ are massless vectors that are orthogonal to $K_1$ and $K_2$ and normalised such that $n\cdot \overline{n} = K^{\flat}_1 \cdot K^{\flat}_2 = \frac{1}{2} \gamma_{12}$. Now one can easily solve three cut conditions
\begin{equation}
l^2 = \mu^2 \qquad (l-K_1)^2 = \mu^2 \qquad (l+K_2)^2 = \mu^2 
\end{equation}
where for generality we have kept massive cut propagators. The solution is
\begin{align}
d &=- \frac{\gamma_{12} a b - \mu^2}{c \gamma_{12}} \\
a & = \frac{K_2^2 (K_1^2 + \gamma_{12})}{K_1^2 K_2^2 - \gamma^2_{12}} \\
b & = - \frac{K_1^2 (K_2^2 + \gamma_{12})}{K_1^2 K_2^2 - \gamma^2_{12}} 
\end{align}
For a quadruply cut propagator,  the final constraint reads 
\begin{equation}
(l+K_2 + K_3)^2 = \mu^2
\end{equation}
This can be written as
\begin{equation}\label{eq:finaleqquadcut}
K_3^2 + 2 K_3 \cdot (l+K_2) = 0
\end{equation}

Let us first verify directly that the integrand of a one loop $\mathcal{N}=4$ SYM amplitude obeys BCFW on-shell recursion relations.  Consider BCFW shifting legs $n$ and $1$. The integrand of an $n$ point planar amplitude at one loop in  $\mathcal{N}=4$ is given as a sum over box integrals,
\begin{equation}
I = \sum_{\textrm{box}} c_b I^4_b
\end{equation}
where the sum is over all possible divisions of the ordered, cyclic set $\{1,2,3,\ldots n\}$ into four ordered subsets. Such a division will be referred to as a 'channel'. For every channel, the box integral is the scalar loop integral for which on the four corners the sum over momenta over the subsets in the channel enters. The box coefficient can be computed easily by quadruple cut \cite{Britto:2004nc}, for which the loop momenta are put on-shell. The result is simply the sum over the two solutions to the on-shell constraints of the product of four tree amplitudes, one for each corner of the box integral
\begin{equation}
c_b = \sum_{\textrm{2 solutions}} \sum_{\textrm{pols}}A_1 A_2 A_3 A_4
\end{equation}

After shifting legs $n$ and $1$, the terms in the above sum split into two classes: if legs $n$ and $1$ appear on the same corner of the box integral, they appear on the same tree level scattering amplitude which is known to obey BCFW. Hence the remaining term has legs $n$ and $1$ appearing on adjacent corners of the box integral. In this case the loop momentum has to be taken to be $l+\hat{n}$ on the line between the two corners. However, the on-shell constraints also depend on the shifting parameter, say $z$. From the above kinematic analysis, it is advantageous to pick the corners without shifted legs to define the momenta $K_1$ and $K_2$. Then, three constraints can be solved without involving $z$. Resolving the final constraints shows, that the solution for $c$ in \eqref{l-decomposition} will be finite in the large $z$ limit. 

From equation \eqref{eq:finaleqquadcut} this also implies that the loop momentum leg between the shifted corners of the box integral has a momentum that becomes orthogonal to $q_{1n}$ in the large shift limit. The sum over polarisations in the connecting leg yields
\begin{equation}
\sum_I \xi^I_{\mu} \xi^{I}_{\nu} = \eta^{\mu\nu} 
\end{equation}
up to terms which vanish by on-shell gauge invariance. Hence this contribution is of order $z^0$. The loop momentum leg yields a suppressing $\frac{1}{z}$. This leaves the behaviour of the tree amplitudes.

Note that the shift on the tree amplitudes is not quite the same as a BCFW shift: the momentum in the loop-momentum dependent leg is non-linear in the shift parameter $z$. 
\begin{equation}
\hat{k}_2 \rightarrow k_2 - q z \qquad (l_2(z)+ K_2 + K_3 + q z)
\end{equation}
This shift is such that
\begin{equation}\label{eq:singularityexpose}
q \cdot  (l(z)+ K_2 + K_3 ) = \frac{(l_2 + k_2)^2 }{2 w} = \mathcal{O} \left(\frac{1}{w} \right)
\end{equation}
The behaviour of the tree amplitude under the shift can be analysed with the same methods as used for ordinary BCFW shifts. Here we employ the off-shell method of \cite{Boels:2010nw}. 

First consider gluons in the loop. In contrast to the ordinary shift case, there are no singular graphs in the lightcone gauge defined by the vector $q$, up to a subtlety discussed below. Hence, a generic diagram will mostly contain negative powers of $z$, as many as there are propagators between the shifted legs in an individual Feynman graph. The exception is the terms in the lightcone gauge propagator which scale as $q_{\mu} q_{\nu} z^0$ (see \cite{Boels:2010nw}). These terms only contribute when the $q$'s contract into external momenta. As shown in \cite{Boels:2010nw}, the leading graph at order $z^0$ are the one where the shifted legs contract into a four vertex, together with the lightcone-gauge propagator contribution of a single hard line.

A further subtlety is that there is a singular class of graphs in the limit $w \rightarrow \infty$: the lightcone gauge propagator contains terms
\begin{equation}
\propto \frac{q_{\mu} P_{\nu} + P_{\mu} q_{\nu} }{q \cdot P}
\end{equation}
for a leg with momentum $P$. Only for the trivalent graph where $\hat{k}_2$ and $P$ enter, there will be a $\tilde w$-type numerator in the lightcone gauge propagator. Note that $\hat{k}_2 + P$ is not independent of $z$, since $l_2$ depends on this. The order $w$ part of the graph contains either a momentum contracted directly into the three vertex, or the lightcone gauge choice vector. By inspection of the three vertex, the first possibility vanishes. The second leads to either a metric contraction, or a $q$ contracted into the external polarisation in the anti-symmetric part of the expression, which is sub-leading. Note that local contributions to the scattering amplitude at order $z^0$ must be anti-symmetric in the external legs by Bose symmetry.

The upshot is that the BCFW shift of terms in the box graphs obey \eqref{eq:bcfwshiftint} separately for gluons in the loop: the two corners of the cut box scale like this, the tensor structure of the loop momentum gluon is contracted with a simple metric, and there is one power of suppression coming from the loop propagator. The case of minimally coupled scalars and fermions in the loop proceeds along very similar lines. The main difference is that the order $\tilde z $ contribution is absent.  

The double recursion relations now follow for box coefficients in quite generic theories as well as for the so-called 'massive box' coefficients within the rational terms. For $\mathcal{N}=4$, this completes the computation of the one-loop integrand from tree amplitudes. It would be interesting to further explore the form of the integrand thus obtained. For triangle and bubble coefficients this conclusion of double recursion certainly holds if limits may be interchanged: both of these contributions involve taking a limit of a product of three and two amplitudes respectively. For the triangle and bubble cases, the massless momenta $K^{\flat}_1$ and $K^{\flat}_2$ will both start to depend on the shift parameter, however in such a way that the cut loop momentum remains finite in the large shift limit. If BCFW shift and these limits may be interchanged, the above analysis follows. This certainly requires more analysis.

\section{Rational terms: from helicity equal to MHV}
Pure Yang-Mills theory contains two infinite series of amplitudes which vanish at tree level, but are non-zero at loop level: these are the gluon amplitudes with all helicities equal, or one unequal. These amplitudes vanish in any supersymmetric theory which conserves $U(1)_R$, as is certainly true at the perturbative level. Since these amplitudes vanish at tree level, they cannot diverge either in the UV or in the IR at loop level. More generally, there are terms known as 'rational terms' which cannot be fixed by four dimensional unitarity cuts and which vanish in any supersymmetric theory. The finite amplitudes are amplitudes which only have rational terms in this parlance. Hence, the rational terms and finite amplitudes are the perfect case to see if the proposal above for loop level recursion works in a realistic theory.

Generically, the finite amplitudes can be computed at one loop by using only a complex scalar in the loop. This follows from a decomposition of the matter content circling in the single loop into supersymmetric multiplets, 
\begin{equation}
A^1_{\mathcal{N}=0} = A^1_{\mathcal{N}=4} - 2 A^1_{\mathcal{N}=2} + A^1_{\textrm{scalar}}
\end{equation}
This can be done gauge-invariantly through the background field method. The tree level amplitudes needed to compute the would-be one loop integrand contain in this case therefore the required gluon content as well as a scalar and an anti-scalar. Here we will focus solely on the leading colour single trace contribution to the one loop amplitude. For these the scalar particles are color-adjacent. These amplitudes are actually proportional to the amplitudes where the scalars are in the fundamental representation. 

A subtle but important point here is the dimensionality of the scalar: it should be a scalar in $D$ dimensions. For the amplitudes used to construct the integrand, this means that the particles in the loop are on-shell in $D$ dimensions. The D-dimensional on-shell momentum can be decomposed as:
\begin{equation}
l^D_{\mu} = l^{4}_{\mu} + l^{-2 \epsilon}_{\mu} 
\end{equation}
so that the D-dimensional massless scalar for which $p^2=0$, is massive from a four dimensional perspective,:
\begin{equation}
(l^{4})^2 = (l^{-2 \epsilon})^2 \equiv \mu^2
\end{equation}
In our following discussion, we will use the capital $L$ to denote the momentum component in 4 dimensions, $L_\mu=l^4_\mu$.  Since this is a momentum of a particle in the loop, this momentum must be integrated over. Often the dimension shifting relation in the Appendix of \cite{Bern:1995db}
\begin{equation}\label{eq:dimrelation}
\int \frac{dl^{D}}{(2 \pi)^{D/2}} \mu^2 \left(\phantom{I} \ldots \phantom{I} \right) = \epsilon \int \frac{dl^{D+2}}{(2 \pi)^{D/2 + 1}}  \left(\phantom{I} \ldots \phantom{I} \right)
\end{equation}
is useful for $D= 4-2 \epsilon$.

\subsection{Helicity equal case}
The helicity equal scattering amplitude after integration is known, see \cite{Bern:1993qk} \cite{Mahlon:1993si}, 
\begin{equation}
A^1(g^+ \ldots g^+) = \frac{\sum_{1\leq i_1 \leq i_2 \leq i_3 \leq i_4 \leq n}  \braket{i_1 i_2} \sbraket{i_2 i_3} \braket{i_3 i_4} \sbraket{i_4 i_1}   }{\braket{12} \ldots \braket{n-1,n} \braket{n1}} + \mathcal{O}(\epsilon)
\end{equation}
Note that the fact that this amplitude only has collinear poles follows from the absence of tree level helicity equal and one-helicity unequal amplitudes (except the three point amplitude) by perturbative unitarity. 

For the helicity equal one-loop amplitude (not the integrand!), on-shell recursion was first explored in \cite{Bern:2005hs}. Performing a BCFW shift of particles $n$ and $1$, it is easy to see that the scattering amplitude will not vanish at infinity. From the above expression the residue at infinity can be computed,
\begin{equation}\label{eq:resatinfallplus}
\textrm{Res}_{z=\infty} \left(A_{n}^1(1,2,\ldots n,)  \right) \propto  - \frac{1}{\braket{n2}} \frac{1}{\langle \langle 2, n \rangle \rangle}  \sum_{2<i_2 < i_3 \leq n}  \sbraket{1 i_2 }  \braket{i_2 i_3 }  \sbraket{i_3 1}
\end{equation}
This computation can be found in appendix \ref{app:allplusres}. On-shell recursion relations for the integrand yield the same tree-like factorisation terms as those relations for the integrated expression. The difference is that for the integrand the single cut contribution must still be computed. Hence, the integrated single-cut contribution must match \eqref{eq:resatinfallplus} so that by induction the correct result is produced.

The seed tree amplitude for the integrand is the amplitude with a complex scalar pair and the rest helicity-equal gluons. This amplitude was computed for instance in \cite{Badger:2005zh, Forde:2005ue,Rodrigo:2005eu}. In the most compact, all multiplicity form listed in \cite{Rodrigo:2005eu}, it reads:
\begin{equation}\label{eq:treelevelscalarallplus}
A_{n}(\phi, g_1^+ \ldots g_n^+, \bar{\phi}) = \mu^2  \frac{1}{\braket{12} \ldots \braket{n-1,n}} \frac{[ 1 | \prod_{k=2}^{n-1} (L_{0,k} - {p \!\!\! \slash}_k {p\!\!\! \slash}_{ 0,k-1} ) | n]}{L_{01} \ldots L_{0,n-1}  }
\end{equation}
where $L_{0,i} = (p_{0,k})^2 - \mu^2$ and $p_{0,k} = L +  \sum_{j=1}^k p_j$. As a special case, the five point amplitude with three gluons reads
\begin{equation}\label{eq:treelevelscalarallplus5pt}
A_3(\phi_1, 1^+, 2^+, 3^+, \bar{\phi}_2) = i\mu^2 \frac{[1|{L \!\!\! \slash}_1{p\!\!\! \slash}_{1,2}|3]}{\braket{12}\braket{23}L_{0,1}L_{0,2}}
\end{equation}

\subsubsection*{Result from on-shell recursive integrand}
The steps needed to construct the seed for the integrand in this case are quite straightforward, as the relevant scalar amplitude in equation \eqref{eq:treelevelscalarallplus} is simply finite in all the limits which need to be taken for application of equation \eqref{eq:masterformulasinglecutsYM}. First study the case of three external gluons. In this case, the integrand is given by taking the limit of $l_1 \rightarrow -l_2$ in equation \eqref{eq:treelevelscalarallplus5pt}. For this
\begin{equation}
\lim_{l_1 \rightarrow -l_2} A_3(\phi_1, 1^+, 2^+, 3^+, \bar{\phi}_2) = 0
\end{equation}
holds by momentum conservation. So the first result is that the all plus three point \emph{integrand} vanishes. This neatly dovetails the knowledge that the integrated three plus amplitude vanishes at one loop, by basic dimensional analysis. 

At four points the computation yields the first non-trivial answer. There is no tree-level contribution at four points to the on-shell recursion relation by the vanishing of the three-plus integrand, so the only term that needs to be computed is the single cut contribution. The integrand through the recursion relations reads
\bea\label{eq:oneloopfourplusstart}
I_4 &=& \mu^2 \frac{1}{(L+p_4)^2-\mu^2} \frac{1}{\braket{\hat{1}2}\braket{23} \braket{34}} \nonumber\\
&&\frac{ [1| [(L-p_3)^2-\mu^2 - p\!\!\! \slash_2 (L\!\!\! \slash+p\!\!\! \slash_4+p\!\!\! \slash_1) ]\,[L^2-\mu^2 - p\!\!\! \slash_3(L\!\!\! \slash+p\!\!\! \slash_4 + p\!\!\! \slash_1 + p\!\!\! \slash_2) ] |\hat{4}]   }{ [(L+p_4+p_1)^2-\mu^2]\,[(L-p_3)^2-\mu^2]\, [L^2-\mu^2]}\,\,
\eea
by deformation pair $|\hat 1\rangle= |1\rangle +z|4\rangle$ and $|\hat 4]=|4]-z|1]$. Note that the appearance of shifted momenta is localised in two places. The shift parameter $z$ is determined to be
\begin{equation}
z^* = \frac{(L+p_4)^2-\mu^2 }{2 q \cdot L},\quad \textrm{with} \quad q= |4\rangle [1| 
\end{equation}
by setting the deformed loop momentum $l+\hat p_4$ on-shell with our chosen convention for momentum routing.

In the special case of four point kinematics one can derive the relation
\begin{equation}\label{Kin:4plus}
\frac{\sbraket{3\hat{4}}}{\braket{\hat{1} 2}} = \frac{\sbraket{34}}{\braket{1 2}}
\end{equation} 
In effect, this is a manifestation of the observation that all angle or square bracket spinor products are proportional to each other in four point kinematics. 
The calculational strategy in the four point case is to move as much as possible ${p \!\!\! \slash}_3$ to the right by using the Clifford algebra. Moreover, using momentum conservation 
\begin{equation}
[1| {p \!\!\! \slash}_2 {L \!\!\! \slash} |\hat{4} ] = - [1| ( {p \!\!\! \slash}_4 +  {p \!\!\! \slash}_3)  {L \!\!\! \slash} |\hat{4} ]  = - \sbraket{14} (2 \hat{p}_4 \cdot L + 2 p_3 \cdot L ) +  [1| {L \!\!\! \slash}  {p \!\!\! \slash}_3 |\hat{4} ].
\end{equation}
Using these observations, one obtains from equation \eqref{eq:oneloopfourplusstart} after some straightforward but tedious algebra
\begin{equation}\label{eq:4pt1loopallplus}
I_4 = \mu^4 \frac{\sbraket{12}\sbraket{34} }{\braket{12} \braket{34} } \frac{1}{ (l+p_4)^2 (l+p_4+p_1)^2(l-p_3)^2 l^2}
\end{equation}
which is the known integrand of the scalar contribution to the four point helicity equal amplitude. Note that the cancellations compared to the original scalar amplitude which led to this result find their origin mostly in the observation that the sum of the gluon momenta is zero. This particular four point computation also appears in \cite{gangandme, hanssenthesis}.

\subsubsection*{All multiplicity}
For higher points in the helicity equal case, we will aim to reproduce the contribution of the residue at infinity, equation \eqref{eq:resatinfallplus} from the loop-like contribution to the integrand. The single cut contribution is guaranteed to match by induction. 

After dimension shifting due to the presence of the $\mu^2$ factor in the tree amplitude, one obtains six dimensional integrals, suppressed by an overall $-\epsilon$. The only divergences of these integrals are of the UV type. These divergences can be isolated easily by powercounting: per additional leg, there is one power of loop momentum in the denominator and two powers in the numerator left. 

In the general multipoint case,
\begin{equation} 
\frac{1}{\braket{\hat{1}2}}   = \frac{1}{\braket{n2}} \frac{\langle n| L|1]}{ (L+p_n+ p_{\gamma})^2 }\quad \textrm{with}\quad p_\gamma \equiv \frac{\braket{12}}{\braket{n2}} q  
\end{equation} 
and 
\begin{equation}
|\hat{n}] = |n] + z^* |1]
\end{equation}
with $z^*$ given by
\begin{equation}
z^* = \frac{(L+p_n)^2}{2 q \cdot L},\quad \textrm{with} \quad q= |n\rangle [1| 
\end{equation}
The integrand is given by
\begin{multline}\label{eq:allplusint}
I_{n}(g_1^+ \ldots g_n^+) = \epsilon \frac{1}{(L+p_n)^2 } \left( \frac{1}{\braket{n2}} \frac{\langle n| L|1]}{ (L+p_n+ p_{\gamma})^2 } \right) \\ 
 \frac{1}{\braket{23} \ldots \braket{n-1,n}} \frac{[ 1 | \prod_{k=2}^{n-1} (L_{0,k} - {p \!\!\! \slash}_k {p\!\!\! \slash}_{ 0,k-1} ) | n]}{L_{01} \ldots L_{0,n-1}  } +\\
 \epsilon  \left( \frac{1}{\braket{n2}} \frac{1}{ (L+p_n+ p_{\gamma})^2 } \right) 
 \frac{1}{\braket{23} \ldots \braket{n-1,n}} \frac{[ 1 | \prod_{k=2}^{n-1} (L_{0,k} - {p \!\!\! \slash}_k {p\!\!\! \slash}_{ 0,k-1} ) | 1]}{L_{01} \ldots L_{0,n-1}  }
\end{multline}
in six dimensions. Let us now isolate the UV divergences in $D=6-2 \epsilon$ from this integrand by powercounting. In the denominator, the minimal amount of propagators is two in the second term and three in the first. There is therefore up to a quadratic divergence in the second and a linear one in the second. The first, quadratic divergence trivially cancels since $[1| 1] = 0$. The linear divergence in the first term cancels since all momenta in the denominator of this divergent term except the loop momentum are orthogonal to the shift vector $q$. Hence this linear divergent term, after introducing Feynman parameters and shifting in the standard fashion will give a vanishing contribution. 

The conclusion is that at least one propagator should be left un-cancelled. It is straightforward to verify that for the first term  in equation \eqref{eq:allplusint}, this is the only possibility to generate UV divergent contributions. To see this, use
\begin{equation}\label{eq:divcontrib}
L_{\mu} L_{\nu} \rightarrow \frac{1}{6-2\epsilon} \eta^{\mu\nu} L^2
\end{equation}
to compute the contribution to the helicity equal amplitude,
\begin{equation}\label{eq:contr1}
I_{\textrm{first} } = - \frac{1}{(4 \pi)^3 \braket{n2} {\langle \langle 2, n \rangle \rangle} }\sum_{k=1}^{n-1} [1| p_k |n\rangle [ 1 | n]
\end{equation}

The second term proceeds similarly. Here, one can have two propagator contributions left un-cancelled, while equation \eqref{eq:divcontrib} will yield an additional `propagator' in the numerator. Since only the UV-divergent contribution needs to be calculated, the details of the loop momentum shift do not matter here. The result is:

\begin{equation}\label{eq:contr2}
I_{\textrm{second}} = +  \frac{1}{(4 \pi)^3 \braket{n2} {\langle \langle 2, n \rangle \rangle} }\sum_{2< j<k\leq n-1} [1| p_j |k\rangle [ 1 | k]
\end{equation}
It is now clear that summing the two contributions in equations \eqref{eq:contr1} and  \eqref{eq:contr2} will yield the known residue of the all-plus amplitude in equation \eqref{eq:resatinfallplus}. It can be checked that all constants work out as well. 

This concludes the inductive step. At four points, we have explicitly demonstrated the corrected-ness of the on-shell recursive integrand in the helicity equal case. At higher point, the single cut terms will contribute a finite piece, which when taken together with the tree-pole factorisations will give the full result, up to order $\epsilon$ terms. 

\subsubsection*{Double cut}
Using the double BCFW-shift construction, the just derived result becomes again a consequence of unitarity for all multiplicity. The single cut of the known answer for the integrand certainly obeys the double BCFW shift-derived secondary recursion relation, as can be checked explicitly from the form given in \cite{Elvang:2011ub}. Then, the terms in this secondary recursive sum are of the double unitarity cut type. Since the form of the integrand written above also obeys on-shell recursion relations and tree-level unitarity to match the double cuts, the two should be the same, up to terms which vanish after integration\footnote{Since a double cut does not localise the $4-2 \epsilon$ dimensional integration completely, the remaining 2 body Lorentz-invariant phase space integral can yield integrates-to-zero type combinations}. 

\subsubsection*{Connection to MHV superamplitudes}
There is an interesting connection between MHV superamplitudes in $\mathcal{N}=4$ super-Yang-Mills theory and the helicity equal amplitudes in pure Yang-Mills at one loop. The latter follows from the former by removing the supermomentum-conserving delta function and replacing it by a $\mu^4$ term on the level of the integrand \cite{Bern:1996ja}. Hence, it should not be a surprise on-shell recursion works for the integrand of the helicity equal terms, as it does for the maximally supersymmetric theory in four dimensions.

\subsection{One helicity unequal}
Let us move forward to one helicity unequal examples. As is true in the general case, the seed tree amplitude diverges in the limit $l_1\rightarrow -l_2$ to obtain a non-singular form $\tilde I_{\textrm{ll}}$. 
Instead of the general residue prescription for the forward limit discussed above, here we employ a more direct strategy. 

The momenta $L_1$ and $L_2$ of complex scalars $\phi_1$ and $\phi_2$ can be written as 
\bea \label{deform1}
L_1^\mu= L^\mu+ {w\over 2}\langle \hat L | \gamma^\mu| 1],\quad
L_2^\mu= -L^\mu+ {u\over 2} \langle n| \gamma^\mu| \hat L].
\eea 
with external momenta
\bea \label{deform2}
\hat p_1^\mu=p_1^\mu -{w\over 2} \langle  \hat L | \gamma^\mu| 1], \quad
\hat p_n^\mu=p_n^\mu -{u \over 2} \langle n| \gamma^\mu| \hat L]
\eea
and other momenta $p_2,\cdots,p_{n-1}$ stay unchanged. This is a double BCFW shift. The loop momentum which will be integrated over is taken to be
\bea
&&L_{\alpha\dot \alpha}= \hat L_\alpha \hat L_{\dot \alpha} +\frac{\mu^2}{\langle n|L|1]} |n\rangle [1|,\quad \textrm{with}\quad  \eta=|n\rangle [1|. 
\eea 

The integrand $\tilde{I}_{\textrm{ll}}$ can be achieved by pushing through limits $u\rightarrow 0$ and $w\rightarrow 0$ successively, i.e.,
\bea
\tilde{I}_{\textrm{ll}} = \lim_{w\rightarrow 0} \lim_{u\rightarrow 0} [A_n(L_1, \hat p_1,\cdots,\hat p_n,L_2)+A_n(L_2, \hat p_1,\cdots,\hat p_2,L_1)].
\eea
Once $u\rightarrow 0$, $L_1\, {\mathbin{\!/\mkern-5mu/\!} }\, L_2$ is realized from \eqref{deform1}; then $w\rightarrow 0$ makes $L_1\rightarrow -L_2$ eventually, in other words, this limit gives the residue of \eqref{softlimit} at $w=0$. 

The new part in this subsection is that the divergent graphs in the limits taken will simply be isolated and dropped. 

\subsubsection{Bubble construction}\label{bubble-oneminus}
\begin{figure}[!htb]
\centering
\includegraphics[scale=0.7]{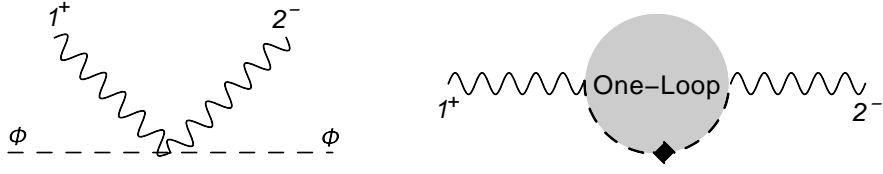}
\caption{Construct $\tilde{I}_{\textrm{ll}, 2}$ from tree-level amplitude, the grey blob on the right represents a tadpole or a bubble.}
\label{fig:bubble-1minus}
\end{figure}

Consider the bubble structure coming from $A_2(\phi_1, 1^+,2^-,\overline \phi_2 )$ \cite{Forde:2005ue} 
\bea
A_2(\phi_1, 1^+,2^-,\overline \phi_2 )=- i {\langle 2|L\!\!\!\slash_1|1]\over (p_1+p_2)^2 [(L_1+p_1)^2-\mu^2]},
\eea
and apply the construction strategy
\bea
\tilde{I}_{\textrm{ll}, 2} = \lim_{w\rightarrow 0} \lim_{u\rightarrow 0} [A_2(L_1, \hat p_1,\hat p_2,L_2)+A_2(L_2, \hat p_1,\hat p_2,L_1)]=i.
\eea
which is simply a number. See fig.\ref{fig:bubble-1minus}, the grey blob means that all kinds of one-loop structures can appear, in this case might be a tadpole or a bubble. 
The integrand, $I_2=i/(L^2-\mu^2)=1/l^2$, is of course vanishing after integrating the loop-momentum $\int d^D l$.

\subsubsection{One-loop 3-pt construction}\label{3pt-oneminus}

\begin{figure}[!htb]
\centering
\includegraphics[scale=0.7]{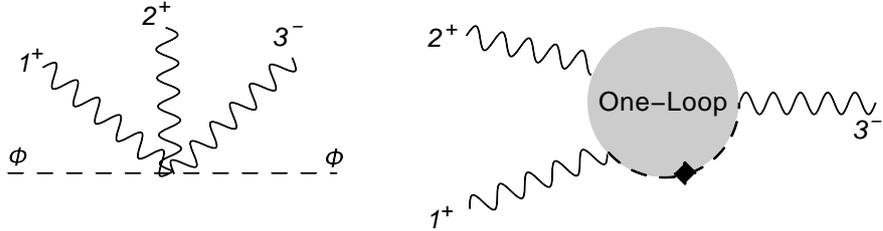}
\caption{Construct $\tilde{I}_{\textrm{ll}, 3}$ from tree-level amplitude. The grey blob means all kinds of loop structures are possibly involved.}
\label{fig:3pt-1minus}
\end{figure}

For three points, one starts from the tree-level amplitude involving two massive complex scalars \cite{Forde:2005ue} ,
\bea\label{tree-3pt}
A_3(\phi_1, 1^+, 2^+, 3^-, \overline \phi_2)&=&-\frac{i\langle3| L\!\!\!\slash_1 (p\!\!\!\slash_1+p\!\!\!\slash_2)|3\rangle^2}{(p_1+p_2+p_3)^2\langle 12\rangle \langle 23\rangle \langle 1| L\!\!\!\slash_1 (p\!\!\!\slash_1+p\!\!\!\slash_2) |3\rangle}\nonumber\\
&&-\frac{i\mu^2\,\langle 2|L\!\!\!\slash_2|3]^2\, [1\,2]}{[(L_1+p_1)^2-\mu^2]\,[(p_3+L_2)^2-\mu^2] [2\, 3]\langle 3| L\!\!\!\slash_2(p\!\!\!\slash_2+p\!\!\!\slash_3 )|1\rangle }.
\eea
Apply the construction strategy, i.e., 1) write $L_1$ and $L_2$ in terms of \eqref{deform1} as well as  $p_1$, $p_3$ and their spinors according to \eqref{deform2}; 2) sum over $A_3(L_1, \hat p_1,p_2, \hat p_3,L_2)$ and $A_3(L_2, \hat p_1, p_2, \hat p_3,L_1)$; 3) take the limits $u\rightarrow 0$ and $w\rightarrow 0$ successively. 
As shown by schematic diagrams fig.\ref{fig:3pt-1minus}, we can derive   
\bea
\tilde{I}_{\textrm{ll}, 3}&=&
=-\frac{i \langle 1 3\rangle}
{ \langle 12\rangle \langle 23\rangle  } 
\frac{\langle3 |L |1]^2}{[(L+p_1)^2-\mu^2]^2}+\frac{i\,\mu^2\,[1\,3]^3}{[1\,2]\,[2\,3]}\frac{1}{[(L+p_1)^2-\mu^2]^2}.
\eea

Consider the loop-level BCFW deformations $|\hat 1\rangle = |1\rangle +z|3\rangle,\,|\hat 3]= |3]-z|1]$ and perform these shifts to construct the integrand
\bea
I_3(z)={i\over (L+\hat p_3)^2-\mu^2}\tilde{I}_{\textrm{ll}, 3}(L\rightarrow L+p_3,z). 
\eea 
Here $\tilde{I}_{\textrm{ll}, 3}(L\rightarrow L+p_3,\, z)$ means the momentum $L$ is first translated to $L+p_3$ and then the BCFW shifts are performed.
The power counting of $z$ shows there is no $z\rightarrow \infty$ boundary contribution here. 
The residue of the pole $z^*$ which satisfies $(L+\hat p_3)^2-\mu^2=0$ is the one-loop 3pt integrand
\bea\label{integrand-3pt-oneminus}
I_3(1^+, 2^+, 3^-)&=&-\text{Res}_{z=z^*} \left(\frac{I_3(z)}{z}\right)=\nonumber\\
&=&-\frac{\langle 1 3\rangle}{ \langle 23\rangle^2} 
\frac{ \langle3 |L |1]^3}{\left[L^2-\mu^2\right]\,\left[(L+p_3)^2-\mu^2\right]\,[(L+p_3+p_1)^2-\mu^2]^2}\nonumber\\
&&-\frac{\mu^2\,[1\,3]^3}{[1\,2]^2}\frac{\langle3 |L |1]}{\left[(L+p_3)^2-\mu^2\right]\,[(L+p_3+p_1)^2-\mu^2]^3}.\,\,
\eea
It is quite straightforward to check the loop-momentum integration over $I_3$ vanishes.

\subsubsection{One-loop 4-pt construction} 

The tree-level amplitude used for 4-pt one-loop integrand construction is\cite{Forde:2005ue}
\bea\label{tree-4pt}
&&A_4(\phi_1, \, 1^+, 2^+, 3^+, 4^-, \,\overline \phi_2)=\nonumber\\
&=& -\frac{i \langle 4|L\!\!\!\slash_1 (p\!\!\!\slash_1+p\!\!\!\slash_2+p\!\!\!\slash_3)|4\rangle^2}{(p_1+p_2+p_3+p_4)^2 \langle 12 \rangle \langle 23 \rangle \langle 34 \rangle \langle 1| L\!\!\!\slash_1 (p\!\!\!\slash_1+p\!\!\!\slash_2+p\!\!\!\slash_3)|4\rangle}\nonumber\\
&&+\frac{i \mu^2 \,\langle 4|L\!\!\!\slash_2 |3]^2 \,\,[3|(p\!\!\!\slash_1+p\!\!\!\slash_2)L\!\!\!\slash_1|1]}{[(L_1+p_1)^2-\mu^2] [(p_3+p_4+L_2)^2-\mu^2][(p_4+L_2)^2-\mu^2] \langle1 2\rangle [3 4] \langle4|L\!\!\!\slash_2 (p\!\!\!\slash_3+p\!\!\!\slash_4)| 2\rangle}\nonumber\\
&&- \frac{i\mu^2 \langle 1 2 \rangle \langle 4|L\!\!\!\slash_2 (p\!\!\!\slash_2+p\!\!\!\slash_3)|4\rangle^2\, \langle 4|(p\!\!\!\slash_2+p\!\!\!\slash_3)|1]}{(p_2+p_3+p_4)^2 [(L_1+p_1)^2-\mu^2] \langle 1 2\rangle \langle 23 \rangle \langle 34 \rangle \langle 4|L\!\!\!\slash_2 (p\!\!\!\slash_2+p\!\!\!\slash_3+p\!\!\!\slash_4)|1\rangle\,\langle 4|L\!\!\!\slash_2 (p\!\!\!\slash_3+p\!\!\!\slash_4)|2\rangle }.
\eea

\begin{figure}[!htb]
\centering
\includegraphics[scale=0.7]{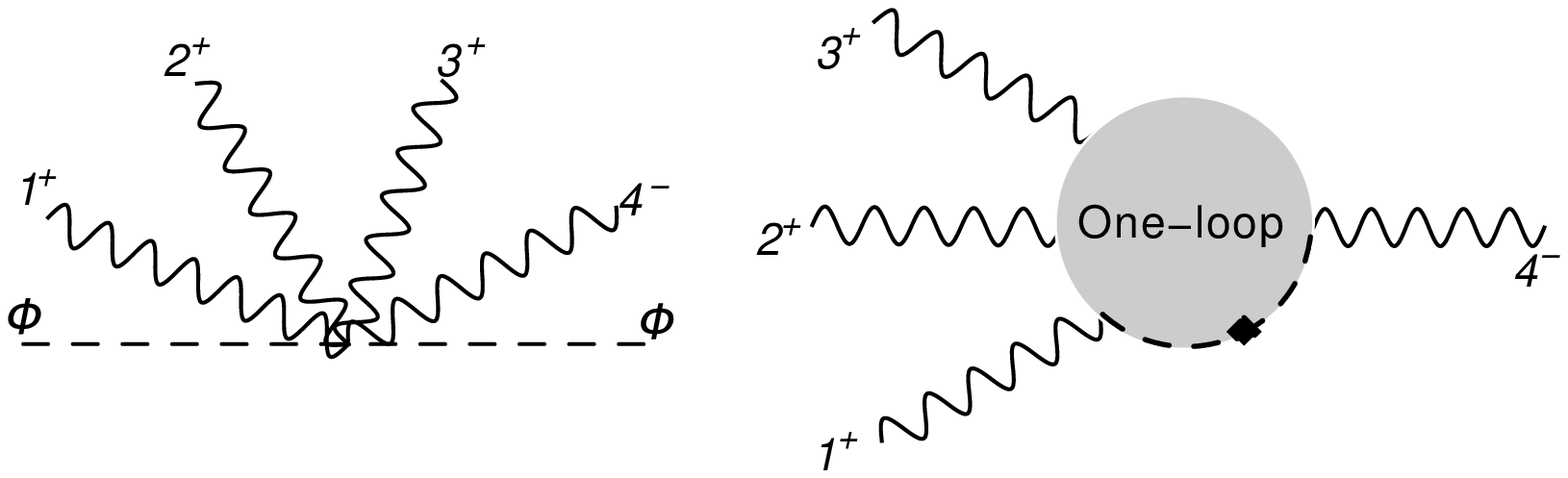}
\caption{Construct $\tilde{I}_{\textrm{ll}, 4}$ from tree-level amplitude}
\label{fig:4pt-1minus}
\end{figure}

Apply the same construction strategy as in 3-pt case, the singularities from collinear and soft limits can be isolated again, see fig.\ref{fig:4pt-1minus} 
\bea
\tilde{I}_{\textrm{ll}, 4}&=&-\frac{i \langle 1 4\rangle}
{ \langle 12\rangle \langle 23\rangle \langle 34\rangle } 
\frac{\langle4 |L |1]^2}{[(L+p_1)^2-\mu^2]^2}\nonumber\\
&&-\frac{i\,2\mu^2}{\langle 1\,2\rangle^2}\frac{\langle 4|L\!\!\!\slash|3]^2}{\left[(L+p_1)^2-\mu^2\right]\,\left[(L+p_1+p_2)^2-\mu^2\right]\,\left[(L+p_1+p_2+p_3)^2-\mu^2\right]}
\eea 

To explore the integrand, we shift the internal momentum $L\rightarrow L+p_4$ in $\tilde{I}_{\textrm{ll}, 4}$ and then impose the the BCFW deformations $|\hat 1\rangle = |1\rangle +z|4\rangle,\,|\hat 4]= |4]-z|1]$. 
The function w.r.t. $z$ is
\bea
I_4(z)={i\over (L+\hat p_4)^2-\mu^2}\tilde{I}_{\textrm{ll}, 4}(L\rightarrow L+p_4,z). 
\eea 
The residue of the pole $z^*$ which satisfies $(L+\hat p_4)^2-\mu^2=0$ is the one-loop 4pt integrand
\bea\label{I40}
I_4&=&-\text{Res}_{z=z^*} \left(\frac{I_4(z)}{z}\right)=\nonumber\\
&=&\frac{\langle 1 4\rangle}
{ \langle4 2\rangle \langle 2 3\rangle \langle 34\rangle } 
{1\over (L+p_4)^2-\mu^2}\frac{\langle4 |L|1]^2}{[(L+p_1+p_4)^2-\mu^2]^2}\frac{\langle4 |L|1]}{(L+p_4+p_{\gamma})^2-\mu^2}\nonumber\\
&&+\frac{2\mu^2}{\langle 4\,2\rangle^2}{1\over (L+ p_4)^2-\mu^2}\frac{\langle 4|L\!\!\!\slash|3]^2}{\left[(L+p_1+p_4)^2-\mu^2\right]\,\left[(L-p_3)^2-\mu^2\right]\,\left[L^2-\mu^2\right]}\frac{\langle4 |L|1]^2}{\left[(L+p_4+p_{\gamma})^2-\mu^2\right]^2}.\nonumber\\
\eea
To make sure our construction, we compute the loop-momentum integration over $I_4$. 
The first line vanish in the loop-momentum integration due to the scale-less denominate, the integration result comes from the second line is
\bea
&&A^1_4(1^+,2^+,3^+,4^-)=-\frac{2i [13]^2}{[41]\langle 12\rangle\langle 23\rangle[34]} {1\over (4\pi)^{2-\epsilon}}  {st\over u }\nonumber\\
&&\qquad \qquad\qquad \left[{t(u-s)\over us} J_3(s) +{s(u-t)\over tu} J_3(t)-{t-u\over s^2} J_2(s)- {s-u\over t^2}\,J_2(t)+{st\over 2u}\,J_4+K_4\right],\nonumber\\
\eea
which matches the known result in \cite{Bern:1995db}, see also \cite{Elvang:2011ub}.

\subsubsection*{Double cut}
Using the double BCFW-shift construction, the just derived result becomes again a consequence of unitarity for all multiplicity, just as it was for the helicity equal case. The single cut of the known answer certainly obeys the double BCFW shift-derived secondary recursion relation, as can be checked explicitly from the form given in \cite{Elvang:2011ub}. Then, the terms in this secondary recursive sum are of the double unitarity cut type. Since the form of the integrand written above also obeys on-shell recursion relations and tree-level unitarity to match the double cuts, the two should be the same, up to terms which vanish after integration\footnote{Since a double cut does not localise the $4-2 \epsilon$ dimensional integration completely, the remaining Lorentz-invariant phase space integral can yield integrates-to-zero type combinations}.

\subsection{MHV rational terms}\label{MHV-main}
The final example is the four point MHV one-loop integrand construction. Again, we start from the tree-level amplitudes $A_6(\phi_1,2^+,3^-,4^-,1^+,\overline\phi_2)$, see \eqref{tree-4pt-MHV}, and choose a slightly different construction of $L_1$ and $L_2$, namely $\eta=|4\rangle [1|$ which is formed by two adjacent external momentum. 

In applying the construction strategy to manipulate the collinear and soft singularities, there is one term whose singularity cannot be dealt with perfectly. Trace back to the origin of this special term, which sits in the third line of \eqref{tree-4pt-MHV}, it comes from the following channel of BCFW recursion relations at tree-level
\bea
A_4^{\textrm{tree}}(L_1,\hat 2^+, \hat P^-,  L_2) {i\over s_{341}} A_4^{\textrm{tree}}(-\hat P^+,\hat 3^-,4^- , 1^+).
\eea
This singularity again comes from the bubble or tadpole like loop structure, which must be amputated in the computation. Since its contribution was argued to integrate to zero above, this term can simply be dropped. 

\begin{figure}[!htb]
\centering
\includegraphics[scale=0.7]{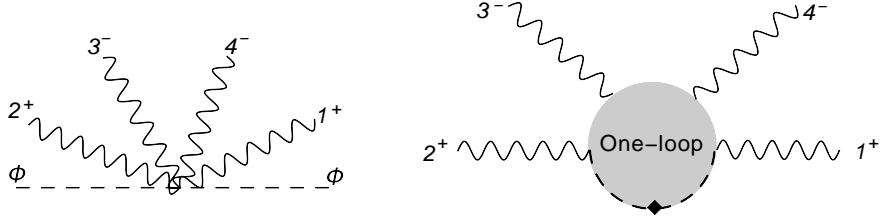}
\caption{Construct $\tilde{I}_{\textrm{ll}, \textrm{MHV}}$ from tree-level amplitude}
\label{fig:MHV-L}
\end{figure}

According to Appendix \ref{MHV-limits}, the non-singular terms turn out $\tilde{I}_{\textrm{ll}, \textrm{MHV}}=\tilde{I}_{\textrm{ll}, T_1}+\tilde{I}_{\textrm{ll}, T_2}+\tilde{I}_{\textrm{ll}, T_4+T_5}$, see the schematic diagram fig.\ref{fig:MHV-L}.

Performing the BCFW deformations $|\hat 1]=|1]-z|2],\,\, |\hat 2\rangle =|2\rangle +z|1\rangle$, then in MHV case, factorized channels possibly appear at tree-like propagators or loop-like propagators. It can be verified that tree-like factorisations vanish as expected, see Appendix \ref{MHV-TL} for detailed computations. The contribution coming from loop-like factorization is 
\bea
I_{\textrm{MHV}}(z)={i\over (L+\widehat p_1)^2-\mu^2} \tilde{I}_{\textrm{ll,\,MHV}}(L\rightarrow L+p_1,z)
\eea
where the internal momentum $L\rightarrow L+p_1$ before applying BCFW deformations.

The residue of $\int dz\,( I_{\textrm{MHV}}(z)/z)$ at $z^*=((L+p_1)^2-\mu^2)/\langle 1|L|2]$ from $(L+\widehat p_1)^2-\mu^2=0$ evaluates to a quite long expression.  After considerable computation, most parts vanish while integrating over the loop-momentum and only three non-vanishing terms are left, see \eqref{integrand_MHV_T1} in Appendix \ref{MHV-Loop}. 
The resulting expression of the MHV integrand which does not vanish after integration is
\bea\label{integrandMHV}
I_{\rm{MHV}}&=&
-\frac{2\,i\,\mu^4\,\langle 3|L+1|2]\,\langle 34\rangle^2}{\langle1 2\rangle\, \langle1 3\rangle\,[(L+p_1+p_\gamma)^2-\mu^2]^2\,[(L+p_1)^2-\mu^2][(L-p_4)^2-\mu^2][L^2-\mu^2]}\nonumber\\
&&+\frac{2\,i\,\mu^4\,\langle 34\rangle^2}{ \langle1 2\rangle^2\,[(L+p_1+p_\gamma)^2-\mu^2]\,[(L+p_1)^2-\mu^2]\,[(L-p_4)^2-\mu^2][L^2-\mu^2]}\nonumber\\
&&-\frac{2\,i\,\mu^4\,\langle 34\rangle^2}{ \langle1 2\rangle^2\,[(L+p_1)^2-\mu^2]\,[(L+p_1+p_2)^2-\mu^2][(L-p_4)^2-\mu^2][L^2-\mu^2]}.\nonumber\\
\eea

To check this form of the integrand, one can compare the integral over this integrand with a known result in  \cite{Bern:1995db}. The symbols $I_2^{D=6-2\epsilon}(t)$, $J_2(t)$ and $K_4$ in following expressions of the loop-momentum integrations are defined in Appendix A of \cite{Bern:1995db}.
\begin{itemize}
\item Denote the first term of $I_{\rm{MHV}}$ in \eqref{integrandMHV} as $I_{\rm{MHV},1}$.
The result of the loop-momentum integration is
\bea
\mathcal I_{\rm{MHV},1}&=&
2 {1\over (4\pi)^{2-\epsilon}} A^{\rm{tree}}(1^+,2^+,3^-,4^-) {1\over t} {1\over (-s_{23})^{-1+\epsilon}}{\Gamma(1+\epsilon) \Gamma(-\epsilon) \Gamma(2-\epsilon)\over \Gamma(4-2\epsilon)}\nonumber\\
&=&2 {1\over (4\pi)^{2-\epsilon}} A^{\rm{tree}}(1^+,2^+,3^-,4^-) {1\over t}  I_2^{D=6-2\epsilon}(t).
\eea
From the first line to the second line Euler's reflection formula $\Gamma(1-z) \Gamma(z) = {\pi\over\sin(\pi z)}$ for $z\in \!\!\!\!\!\slash \,\, {\bf Z}$ was used.


\item Denote the second term of $I_{\rm{MHV}}$ in \eqref{integrandMHV} as $I_{\rm{MHV},2}$.
The integral over the loop-momentum is 
\bea
\mathcal I_{\rm{MHV},2}&=&-{2\langle 34\rangle^2 \over \langle 12\rangle^2 }{1\over (4\pi)^{2-\epsilon}}(-\epsilon) {1\over (-s_{23})^{\epsilon}} {\Gamma(\epsilon)\Gamma(1-\epsilon)\Gamma(2-\epsilon) \over \Gamma(4-2\epsilon)}\nonumber\\
&=&2 {1\over (4\pi)^{2-\epsilon}} A^{\rm{tree}}(1^+,2^+,3^-,4^-) {1\over s}  J_2(t).
\eea
Again, Euler's reflection formula was applied.


\item The second term of $I_{\rm{MHV}}$ in \eqref{integrandMHV} is denoted as $I_{\rm{MHV},3}$.
The result of the loop-momentum integration is 
\bea
\mathcal I_{\rm{MHV},3}&=&2 A^{\rm{tree}}(1^+,2^+,3^-,4^-)\left(-{t\over s} K_4\right) 
\eea

\end{itemize}

Summing $\mathcal I_{\rm{MHV},1}$, $\mathcal I_{\rm{MHV},2}$ and $\mathcal I_{\rm{MHV},3}$ together, 
\bea
A_4^{\rm{L, scalar}}(1^+,2^+,3^-,4^-) ={1\over (4\pi)^{2-\epsilon}} A_4^{\rm{tree}} \left({1\over t} I_2^{D=6-2\epsilon}+{1\over s} J_2(t) -{t\over s}K_4\right).
\eea
is obtained, which is indeed the known expression.

\section{Gravity four point amplitudes: all-plus and maximal SUSY}

In this section, we illustrate our idea about the double recursion relations by two further concrete examples: the gravity all-plus four point amplitude as well as its closely related cousin, the maximally supersymmetric $\mathcal{N}=8$ amplitude, see \cite{Bern:1996ja}. First, we construct the double cut contribution by multiplying two tree amplitudes. Then a momentum routing is invented such that the on-shell recursive formula reproduces the known result. 

\subsubsection*{all-plus four point gravity amplitude}

In this case, there exist three double-cut contributions, from $s_{12}$-cut, $s_{13}$-cut and $s_{14}$-cut respectively\cite{Bern:1998sv}.  

\begin{figure}[!htb]
\centering
\includegraphics[scale=0.8]{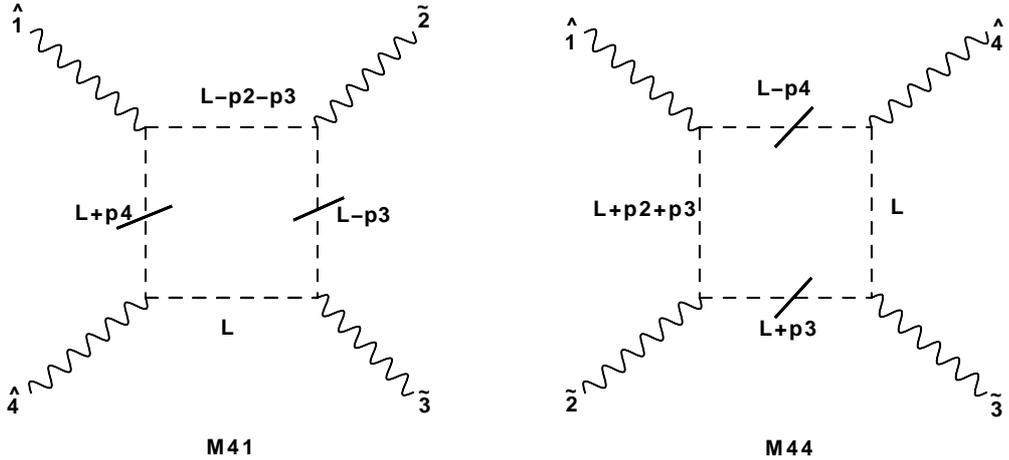}
\caption{This figure contains two contributions from $M_{41}$ and $M_{44}$. The first recursion relation by shifting $1$ and $4$ is equivalent to the single cut on the loop leg between $1$ and $4$. The second recursion relation by shifting $2$ and $3$ singles out the other cut leg between $2$ and $3$. }
\label{fig:graviton-s12-1}
\end{figure}

\begin{figure}[!htb]
\centering
\includegraphics[scale=0.8]{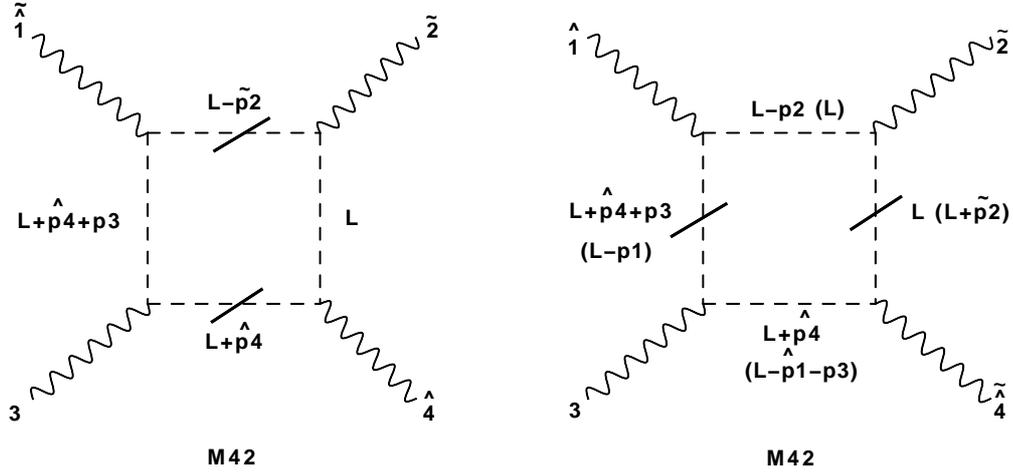}
\caption{The first recursion relations arise from two single poles in $M_{42}$ under BCFW shifts $1$ and $4$, $(L+\hat p_4)^2-\mu^2=0$ in the left digram and $(L+\hat p_4+p_3)^2-\mu^2=0$ in the right. In the left digram, the second recursion relations work under the BCFW shifts on $| \hat 1\rangle $ and $|2]$. In the right diagram, the loop momentum has to be translated as $L\rightarrow L+p_2$ first and then perform the second recursion relations with BCFW shifts on $|2\rangle$ and $|\hat 4]$.}
\label{fig:graviton-s12-2}
\end{figure}

\begin{figure}[!htb]
\centering
\includegraphics[scale=0.8]{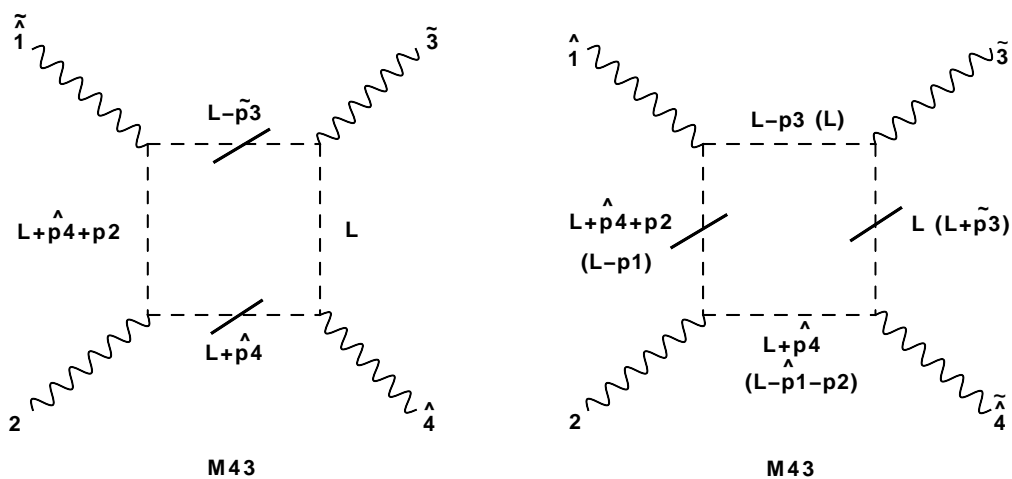}
\caption{The first recursion relations arise from two single poles in $M_{43}$ under BCFW shifts $1$ and $4$, $(L+\hat p_4)^2-\mu^2=0$ in the left digram and $(L+\hat p_4+p_2)^2-\mu^2=0$ in the right. In the left digram, the second recursion relations work under the BCFW shifts on $| \hat 1\rangle $ and $|3]$. In the right digram, the loop momentum has to be translated as $L\rightarrow L+p_3$ first and then perform the second recursion relations with BCFW shifts on $|3\rangle$ and $|\hat 4]$.}
\label{fig:graviton-s12-3}
\end{figure}

Start from $s_{12}$-cut contribution. 
It is derived from connecting two tree-level amplitudes $M_4(\phi_1,1^+,2^+,\overline \phi_2)$ and $M_4(\phi_2,3^+,4^+,\overline \phi_1)$, which can be derived by KLT relations from gauge field amplitudes \cite{Kawai:1985xq, Bern:1998sv}, as   
\bea\label{prel_1}
 M_4(1^+,2^+,3^+,4^+)&=&{i\over L_1^2-\mu^2} {i\over L_3^2-\mu^2}  M_4^{\rm{tree}}(-L_1^s,1^+,2^+,L_3^s)\, M_4^{\rm{tree}}(-L_3^s,3^+,4^+,L_1^s)\nonumber\\
 &=&\mu^8{[12]^2 [34]^2\over \langle 12\rangle^2 \langle 34\rangle^2} {1\over L^2-\mu^2}\left[{1\over(L+p_1)^2-\mu^2} +{1\over(L+p_2)^2-\mu^2}\right]\nonumber\\
&&\quad {1\over (L+p_1+p_2)^2-\mu^2}\left[{1\over(L-p_4)^2-\mu^2} +{1\over(L-p_3)^2-\mu^2}\right]. 
\eea
where $L_1=L$, $L_3=L+p_1+p_2$ in 4-Dim.
The result in \eqref{prel_1} contains four box-type integrand-like ingredients corresponding to external legs of orderings $(1^+,2^+,3^+,4^+)$, $(1^+,2^+,4^+,3^+)$, $(1^+,3^+,4^+,2^+)$ and $(1^+,4^+,3^+,2^+)$ respectively. 
As the loop-momentum $L$ can be shifted arbitrarily, the conventions for later discussion are convenient to be set as shown in the figure \ref{fig:graviton-s12-1}, \ref{fig:graviton-s12-2} and  \ref{fig:graviton-s12-3}, where the integrand-like ingredients in \eqref{prel_1} are written out explicitly, after loop-momentum adjusting, as
\bea
M_{41}&=&\mu^8{[12]^2 [34]^2\over \langle 12\rangle^2 \langle 34\rangle^2} {1\over ((L+p_4)^2-\mu^2)(L^2-\mu^2)((L-p_2-p_3)^2-\mu^2)((L-p_3)^2-\mu^2)},\quad \label{M_41}\\
M_{42}&=&\mu^8{[12]^2 [34]^2\over \langle 12\rangle^2 \langle 34\rangle^2} {1\over((L+p_4)^2-\mu^2) ((L+p_4+p_3)^2-\mu^2)((L-p_2)^2-\mu^2)(L^2-\mu^2)},\quad\label{M_42}\\
M_{43}&=&\mu^8{[12]^2 [34]^2\over \langle 12\rangle^2 \langle 34\rangle^2} {1\over ((L+p_4)^2-\mu^2)((L+p_4+p_2)^2-\mu^2)((L-p_3)^2-\mu^2)(L^2-\mu^2)},\quad \label{M_43}\\
M_{44}&=&\mu^8{[12]^2 [34]^2\over \langle 12\rangle^2 \langle 34\rangle^2} {1\over ((L-p_4)-\mu^2)(L^2-\mu^2)((L+p_3)^2-\mu^2)((L+p_2+p_3)^2-\mu^2)}.\quad \label{M_44}
\eea 

For the first recursion relation, perform the BCFW shifts on $|1\rangle $ and $|4]$, i.e. $|\hat1\rangle=|1\rangle+z|4\rangle$ and $|\hat 4]=|4]-z|1]$ with $q^\mu=\langle 4|\gamma^\mu|1]/2$. 
According to \eqref{Kin:4plus}, the pre-factors in \eqref{M_41}-\eqref{M_44} does not include any parameter $z$.

See figure \ref{fig:graviton-s12-1}, in $M_{41}(z)$ and $M_{44}(z)$, only one singularity appears in each of them, namely $(L+\widehat p_4)^2-\mu^2=0$ and $(L-\widehat p_4)^2-\mu^2=0$ respectively. 
Their residues turn out the same as \eqref{M_41} and \eqref{M_44}.
Then it's natural to perform the second recursion relation by BCFW shifting $2$ and $3$ for results from $M_{41}(z)$ and $M_{44}(z)$.
This time, the poles $(L-\widetilde p_3)^2-\mu^2=0$ and $(L+\widetilde p_3)^2-\mu^2=0$ are selected, which figure out the cut legs opposite to the ones in the first recursion.
Double recursion relations working in $M_{41}$ and $M_{44}$ will contribute to the unitarity $s$-channel cuts in the master integrals. 

In $M_{42}(z)$ and $M_{43}(z)$, each of them contains two single poles, in other words, two possible single cut ways in each of them.
Take $M_{42}(z)$ as an example, BCFW shifts $1$ and $4$ give
\bea
M_{42}(z)&=&\mu^8{[12]^2 [34]^2\over \langle 12\rangle^2 \langle 34\rangle^2} {1\over((L+\hat p_4)^2-\mu^2) ((L+\hat p_4+p_3)^2-\mu^2)((L-p_2)^2-\mu^2)(L^2-\mu^2)}.\qquad
\eea 

\begin{itemize}
\item
Consider the pole $(L+\hat p_4)^2-\mu^2=0$, the solution $z$ is
\bea
z_1^*={(L+p_4)^2-\mu^2\over \langle 4|L|1]}.
\eea
Then the residue of $z_1^*$ gives 
\bea\label{FirstRecursion:M42-1}
{\rm M}_{42}^{(1)}&=&\mu^8{[12]^2 [3 4]^2\over \langle 12\rangle^2 \langle 34\rangle^2} {1\over((L+p_4)^2-\mu^2) ((L-p_2)^2-\mu^2)(L^2-\mu^2)}\nonumber\\
&&\qquad \times {q\cdot L\over 2 (q\cdot L) (p_3\cdot(L+p_4))-(q\cdot p_3) [(L+p_4)^2-\mu^2]}
\eea
This pole, from the previous discussion, is equivalent to the single cut on the same loop-leg.

For the second recursion relation, perform BCFW shifts on $|\hat 1\rangle$ and $|2]$ in \eqref{FirstRecursion:M42-1}, i.e. $|\widetilde {\hat1}\rangle \rightarrow |\hat 1\rangle +z^\prime|2\rangle $ and $|\widetilde 2]\rightarrow |2]-z^\prime |1]$, then the singularity comes from $(L-\widetilde p_2)^2-\mu^2=0$. This pole singularity is again equivalent to a cut.
The residue turns out to keep \eqref{FirstRecursion:M42-1} unchanged.
Note that this double recursion contributes to the unitarity $t$-channel cuts in the master integrals.

\item 
Consider the pole $(L+\hat p_4+p_3)^2-\mu^2=0$, the solution $z$ is
\bea
z_2^*={(L+p_4+p_3)^2-\mu^2\over \langle 4|L+p_3|1]}.
\eea
Then the residue of $z_2^*$ gives 
\bea\label{FirstRecursion:M42-2}
{\rm M}_{42}^{(2)}&=&\mu^8{[12]^2 [3 4]^2\over \langle 12\rangle^2 \langle 34\rangle^2} {1\over((L+p_4+p_3)^2-\mu^2) ((L-p_2)^2-\mu^2)(L^2-\mu^2)}\nonumber\\
&&\qquad \times {-q\cdot (L+p_3)\over 2 (q\cdot L) (p_3\cdot(L+p_4))-(q\cdot p_3) [(L+p_4)^2-\mu^2]}
\eea
Again, this pole is equivalent to the single cut on the same loop-leg.

The second recursion relation is performed after a loop momentum translation, but not directly applied as in \eqref{FirstRecursion:M42-1}. 
The loop momentum in \eqref{FirstRecursion:M42-2} translates as $L\rightarrow L+p_2$, see figure \ref{fig:graviton-s12-2}. 
Then the starting formula of the second recursion becomes 
\bea\label{FirstRecursion:M42-3}
{\rm M}_{42}^{(3)}&=&\mu^8{[12]^2 [3 4]^2\over \langle 12\rangle^2 \langle 34\rangle^2} {1\over((L-p_1)^2-\mu^2) ((L+p_2)^2-\mu^2)(L^2-\mu^2)}\nonumber\\
&&\qquad \times {-q\cdot L\over 2 (q\cdot (L-p_1)) (p_3\cdot(L-p_1))-(q\cdot p_3) [(L-p_1-p_3)^2-\mu^2]}
\eea
As shown on the right diagram in figure \ref{fig:graviton-s12-2}, with the loop momentum settings in the parentheses, one can perform BCFW shifts for $|2\rangle$ and $|\hat4]$, i.e. $|\widetilde 2\rangle=|2\rangle+z^\prime |4\rangle$ and $|\widetilde{\hat 4}]=|\hat 4]-z^\prime |2]$, then the singularity comes from $(L+\widetilde p_2)^2-\mu^2=0$. 
This pole is again equivalent to a cut.
The residue turns out to keep \eqref{FirstRecursion:M42-3} unchanged.
Note that this double recursion contributes to the unitarity $s$-channel cuts in the master integrals.
\end{itemize}

The double recursion relations in $M_{43}$ can be derived in a similar way as $M_{42}$, see the figure \ref{fig:graviton-s12-3}.
Those two results from double recursion in the left and right diagrams contribute to $s$-channel and $t$-channel cuts in the master integrals respectively.

The other two contributions from $s_{13}$ and $s_{14}$ type\cite{Bern:1998sv} can be applied double recursion relations to figure out the cuts in the master integrals and the coefficients of them along the same way for $s_{12}$-cut $M_4(1^+,2^+,3^+,4^+)$.

Sum all terms from double recursion, the integrand is the same as the known one 
\bea
&&I(+,+,+,+)=4\mu^8{[12]^2 [34]^2\over \langle 12\rangle^2 \langle 34\rangle^2}(I^{1234}+I^{1243}+I^{1423})
\eea 
up to terms which vanish after integration, where 
\bea
I^{1234}={1\over (L^2-\mu^2)[(L+p_1)^2-\mu^2][(L+p_1+p_2)^2-\mu^2][(L-p_4)^2-\mu^2]}.
\eea
is basically the integrand of the massless box integral.

\subsection*{Extension to maximally supersymmetric four point graviton amplitude}
The maximal four point supergravity computation is in essence completely similar. The product of two four point tree amplitudes in higher dimensional spinor helicity follows exactly as the result above, see the relevant appendix in \cite{Boels:2012ie}. The BCFW shifts leave the supersymmetric, completely symmetric four point tree-like factor which follows from the supersymmetric delta function $\delta^{16}(Q)$ unchanged, and the computation is from this point forward exactly the same, up to the different pre-factor, in line with \cite{Bern:1996ja}. We note that this basically uses the so-called rung-rule to calculate one loop amplitudes. An extension of this observation to higher points would obviously be very interesting, but might be hard because of the special four point kinematics.

\section{Discussion and conclusions}

In this article we have introduced an explicit definition of the so-called single cut terms which is needed to complete on-shell recursion relations at loop level. We have exhibited general examples and given arguments for the correctness of the proposal, both from a Feynman graph point of view as well as an explicit analysis of the integrand of one-loop amplitudes. Several worked-out examples showcase the method. Unfortunately, they also showcase some of the drawbacks: without further work, the resulting integrands appear hard to integrate directly. Moreover, they contain spurious singularities which vanish after integration. This is a common feature of most recent developments of new calculational technology at the loop level, and is the most pressing obstacle to be overcome. We note that for maximally supersymmetric gauge theories, progress in this direction has been achieved, see e.g. \cite{ArkaniHamed:2010gh} \cite{Bourjaily:2015jna}. A particular goal is to generalise recent progress for helicity equal amplitudes at the two loop level \cite{Dunbar:2016cxp}, \cite{Badger:2016ozq}. The structure of the soft singularities in the limit where the two would-be loop momentum legs have exactly opposite momenta deserves further study. It seems likely that there is a relation to recently studied sub-leading soft divergences in Yang-Mills theories \cite{Casali:2014xpa} (see also \cite{Low:1954kd}). Certainly, the divergences should be amenable to techniques explored in \cite{Bern:2014vva}.

Our results have potential applications beyond standard on-shell recursion. Within the loop-tree duality method introduced in  \cite{Catani:2008xa} for instance, our proposal provides a potentially powerful shortcut to compute complete scattering amplitudes at one loop. This might have interesting applications for the cancellations of divergences in the full cross-section. This has been recently implemented in this method already in a more direct way in \cite{Sborlini:2016gbr}, and it would be interesting to see if this could be generalised. Cancellation of divergences especially beyond the one-loop order is a major concern which has received more attention recently, see for instance \cite{Seth:2016hmv}. We strongly suspect that the loop-tree duality method may be obtained from a special form of the so-called all-line shift  \cite{Elvang:2013cua}. This has the potential to considerably speed up applications of the loop-tree duality method. 

A proof of our proposal for the completion of the on-shell recursion relations at generic loop orders beyond one loop order would be very welcome. In addition, any applications of the developed technology to concrete scattering amplitudes beyond the one loop level would be interesting - this is after all the current frontier of development. Exploring double shifts beyond the one loop order might prove especially fruitful in this context. 

\acknowledgments
It is a pleasure to thank Henrik Hanssen and Gang Yang for collaboration in (very) early stages of this project. RB would like to thank Lionel Mason for a discussion. This work was supported by the German Science Foundation (DFG) within the Collaborative Research Center 676 ``Particles, Strings and the Early Universe".

\appendix
\section{Residue at infinity for helicity equal amplitudes}\label{app:allplusres}
The residue quoted in the text in equation \eqref{eq:resatinfallplus} is straightforward to compute, 
\begin{multline}
\textrm{Res}_{z=\infty} \left(A_{n}^1(1,2,\ldots n,)  \right) \propto \frac{1}{\braket{n2}} \frac{1}{\langle \langle 2, n \rangle \rangle \braket{n 1}} \left( \sum_{1<i_2<i_3<i_4 \leq n} \braket{n i_2} \sbraket{i_2 i_3} \braket{i_3 i_4} \sbraket{i_4 1} \right. \\ - \left. \sum_{1<i_1<i_2<i_3<n} \braket{i_1 i_2} \sbraket{i_2 i_3} \braket{i_3 n} \sbraket{1 i_1}  \right)
\end{multline}
The expression in brackets can be written as
\begin{equation}
\left( \sum_{1<i_1<i_2<i_3 \leq n} \braket{n i_1} \sbraket{i_1 i_2} \braket{i_2 i_3} \sbraket{i_3 1} - \sum_{1<i_3<i_2<i_1<n}  \braket{ n i_1} \sbraket{i_1 i_2 }  \braket{i_2 i_3 }  \sbraket{ i_3 1}  \right)
\end{equation}
which reads
\begin{equation}
\left( \sum_{i_2 = 3}^{n-1} \sum_{i_1=2}^{i_2-1} \sum_{i_3=i_2+1}^{n}  \braket{n i_1} \sbraket{i_1 i_2} \braket{i_2 i_3} \sbraket{i_3 1} - \sum_{i_2 = 3}^{n-2}  \sum_{i_1=i_2+1}^{n-1} \sum_{i_3=2}^{i_2-1}  \braket{ n i_1} \sbraket{i_1 i_2 }  \braket{i_2 i_3 }  \sbraket{ i_3 1}  \right)
\end{equation}
which leads to 
\begin{equation}
 \sum_{i_1=2}^{n-2} \braket{n i_1} \sbraket{i_1 n-1} \braket{n-1 n} \sbraket{n 1} +   \sum_{i_2 = 3}^{n-2}  \left( \sum_{i_1=2}^{i_2-1}  \sum_{i_3=i_2}^{n}  \braket{n i_1} \sbraket{i_1 i_2} \braket{i_2 i_3} \sbraket{i_3 1} -   \sum_{i_1=1}^{i_2} \sum_{i_3=i_2}^{n}  \braket{ n i_1} \sbraket{i_1 i_2 }  \braket{i_2 i_3 }  \sbraket{i_3 1} \right) 
\end{equation}
and from there to:
\begin{equation}
- \braket{n 1} \sbraket{1 n-1} \braket{n-1 n} \sbraket{n 1} -   \sum_{i_2 = 3}^{n-2} \sum_{i_3=i_2}^{n}  \braket{ n 1} \sbraket{1 i_2 }  \braket{i_2 i_3 }  \sbraket{i_3 1}   = -   \sum_{i_2 = 3}^{n-1} \sum_{i_3=i_2+1}^{n}  \braket{ n 1} \sbraket{1 i_2 }  \braket{i_2 i_3 }  \sbraket{i_3 1}
\end{equation}
and hence
\begin{equation}
\textrm{Res}_{z=\infty} \left(A_{n}^1(1,2,\ldots n,)  \right) \propto  - \frac{1}{\braket{n2}} \frac{1}{\langle \langle 2, n \rangle \rangle}  \sum_{2<i_2 < i_3 \leq n}  \sbraket{1 i_2 }  \braket{i_2 i_3 }  \sbraket{i_3 1}
\end{equation}
This is equation \eqref{eq:resatinfallplus}.

\section{General momentum setup}\label{Setting}
To manipulate the forward limits appearing in \eqref{eq:guess1}, we must properly regulate external momenta and eventually take the collinear and soft limits to form a non-singular $\tilde{I}_{\textrm{ll}}$. 

Note the momentum $l$ in $D=4-2\epsilon$ dimension can be decomposed into a 4-Dim component $L$ and a $(-2\epsilon)$-Dim component $\tilde l$. 
The on-shellness for $l$, namely $l^2=0$, indicates $L^2=\tilde l^2=\mu^2$, where the 4-Dim massive momentum can be further written as
\bea
L=\hat L +{\mu^2\over 2 \eta\cdot L} \eta,
\eea
and $\eta$ and $\hat L$ are all null-momentum. 

In particular, consider two $D$-Dim external momenta, their 4-Dim components $L_1$ and $L_2$ would go as $L_1=-L_2=L$ in the limits.
The momentum regularizations which can keep the mass of $L_1$ and $L_2$ as $\mu^2$ are as following 
\bea
L_1^\mu&=& L^\mu+ {b_1\over2} \langle \eta|\gamma^\mu|\hat L] +{c_1\over 2 } \langle \hat L|\gamma^\mu|\eta]+ b_1 c_1 \eta\\
L_2^\mu&=&-L^\mu+ {b_2\over2} \langle \eta|\gamma^\mu|\hat L] +{c_2\over 2 } \langle \hat L|\gamma^\mu|\eta]-b_2 c_2 \eta.
\eea
One can easily check $L_1^2=L_2^2=\mu^2$. 

The collinear and soft limits result in $(b_1+b_2)(c_1+c_2)=0$. 

In principle, the construction does not depend on the choices of $\eta$, so a specific $\eta$ can be chosen to simplify the calculations.

\begin{figure}[!htb]
\centering
\includegraphics[scale=0.9]{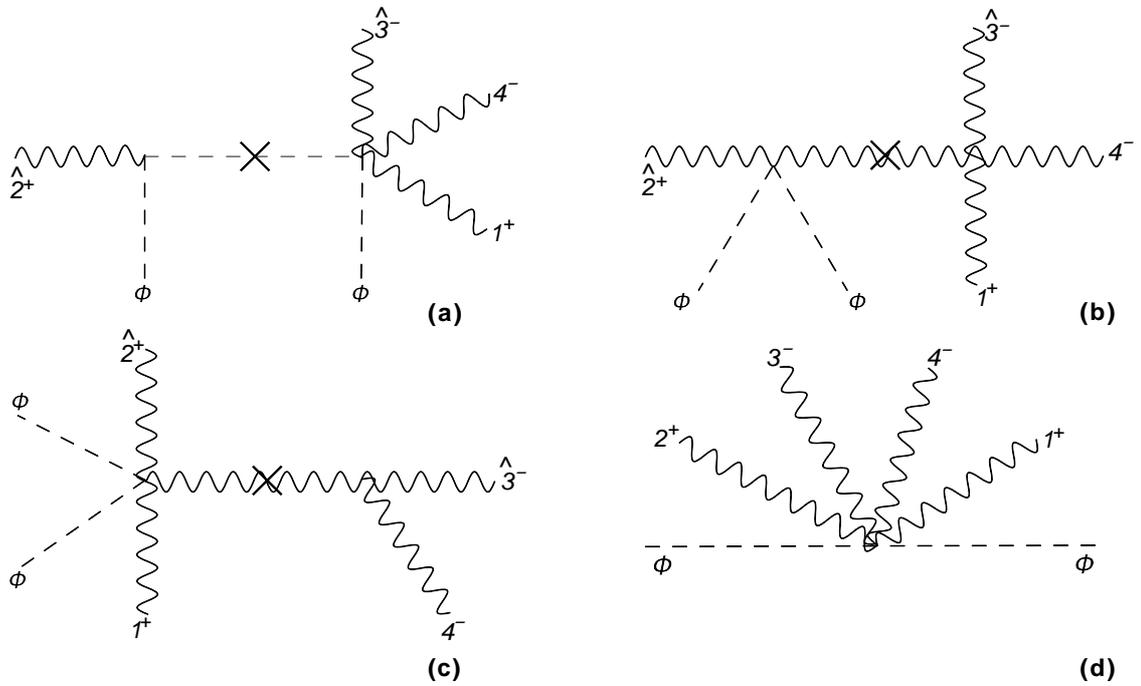}
\caption{MHV tree-level amplitudes}
\label{fig:MHV-tree}
\end{figure}

\section{Details of MHV rational term}\label{MHV} 

One can compute the tree-level 6pt MHV amplitude involving two massive complex scalars by normal BCFW recursion relations \cite{Badger:2005zh}
\bea\label{tree-4pt-MHV}
&&A_6(\phi_1,2^+,3^-,4^-,1^+,\overline \phi_2)=\nonumber\\
&&=-\frac{\langle3|L_1|2 ]^2\,(Q_3\,[1|L_1+p_2|3\rangle-\mu^2\, [14]\, \langle43\rangle)^2}{Q_1\,Q_2\,Q_3\,\langle 23 \rangle \,[41]\, [4|(p_2+p_3) L_1|2]\,\langle 1|L_2 (p_4+p_1)|3\rangle}\nonumber\\
&&-\frac{\mu^2\,\langle 34\rangle^3\,\langle3|L_1 |2]^2 }{Q_1\,\langle 23\rangle\,\langle 41\rangle\,\langle 1|L_2(p_4+p_1)|3\rangle\,\langle3|(p_4+p_1) (L_1+L_2) L_1|2 ]}\nonumber\\
&&+\frac{[2|L_2 L_1|2]^2 \,\langle 34\rangle^3}{s_{1234}\, s_{341}\,\langle 41\rangle\, \langle3|(p_4+p_1) (L_1+L_2)L_1|2] \,\langle1|p_3+p_4|2 ]}\nonumber\\
&&-\frac{[2|(p_3+p_4) L_1|2]^2\, [2|(p_3+p_4) L_2|1]^2}{s_{234}\, Q_3\,[23]\,[34]\,[4|(p_2+p_3)L_1|2]\,[1|(p_2+p_3+p_4) L_1|2]\,\langle 1|p_3+p_4|2]}\nonumber\\
&&+\frac{\mu^2\,[21]^4}{s_{1234}\,[23] \,[34]\,[41]\,[1|(p_2+p_3+p_4) L_1|2]},
\eea
where $Q_1=(L_1+p_2)^2-\mu^2$, $Q_2=(L_1+p_2+p_3)^2-\mu^2$, $Q_3=(L_2+p_1)^2-\mu^2$ and $s_{i_1,\dots,i_m}=(p_1+\dots+p_m)^2$. 
Note that this expression is derived from tree-level BCFW recursion relations through momentum deformation $|\hat 2\rangle = |2\rangle +z|3\rangle ,\,\, |\hat 3] = |3] -z|2]$, as shown in fig.\ref{fig:MHV-tree}. 

\subsection{Manipulation of the forward limits}\label{MHV-limits}
Similar to the unequal momentum regulation, in this MHV case we choose $L_{\alpha\dot \alpha}= \hat L_\alpha \hat L_{\dot \alpha} +\frac{\mu^2}{\langle 1|L|4]} |1\rangle [4|$, the momenta $L_1$ and $L_2$ of complex scalar $\phi_1$ and $\phi_2$ and regulated $p_1$ and $p_4$ can be written as 
\bea 
&&L_1^\mu= L^\mu+ {w\over 2} \langle \hat L | \gamma^\mu| 4],\quad L_2^\mu= -L^\mu+ {u\over 2}\langle 1| \gamma^\mu| \hat L],\nonumber\\
&&\hat p_1=p_1-{u\over 2} \langle 1 | \gamma^\mu| \hat L], \quad \,\,\, \hat p_4=p_4- {w\over 2} \langle \hat L | \gamma^\mu| 4].\nonumber\\
\eea
Again, we hope to gain non-singular expression by summing over permutation of $L_1$ and $L_2$ and taking limits $u\rightarrow 0$ and $w\rightarrow 0$ successively.

For the first line in \eqref{tree-4pt-MHV}, we can obtain
\bea
\tilde{I}_{\textrm{ll}, T_1}
%
&=&\frac{2\,\langle 3|L|2]^2\,[(L-p_1)^2-\mu^2]\,(\langle 3|L|1]-\langle 3|4|1])^2}{\langle 1|L|2]^2 \,\langle2\,3\rangle^2\,[41]^2\,[(L+p_2)^2-\mu^2]\,[(L+p_2+p_3)^2-\mu^2]}\nonumber\\
&-&\frac{4\,\langle 3|L|2]^2\,\mu^2\, (\langle 3|L|1]-\langle 3|4|1])\langle 3 4\rangle}{\langle 1|L|2]^2 \,\langle2\,3\rangle^2\,[41]\,[(L+p_2)^2-\mu^2]\,[(L+p_2+p_3)^2-\mu^2]}\nonumber\\
&+&\frac{2\,\mu^4\,\langle 3|L|2]^2\,\langle 34\rangle^2}{\langle 1|L|2]^2 \,\langle2\,3\rangle^2\,[(L+p_2)^2-\mu^2][(L+p_2+p_3)^2-\mu^2][(L-p_1)^2-\mu^2]}.
\eea
For the first line in \eqref{tree-4pt-MHV}, we can obtain
\bea
\tilde{I}_{\textrm{ll}, T_2}
%
&=&\frac{2\,\langle 34\rangle^3 \,\langle 3|L| 2] \, \langle 3|L|4]}{\langle 14\rangle\, \langle 23\rangle^3\, [42] \, \langle 1|L|2]\,[(L+p_2)^2-\mu^2]}
+\frac{\mu^2 \langle 34\rangle^3 \,\langle 21\rangle\,\langle 3|L| 2]^2 }{\langle 14\rangle\, \langle 23\rangle^3\,\langle 1|L|2]^2 \,[(L+p_2)^2-\mu^2]^2}\nonumber\\
\eea
The third term after manipulation is
\bea
\tilde{I}_{\textrm{ll}, T_3}= O\left ({1\over w}\right),
\eea
which is a divergent part. 
Note that this term comes from the tree-level structure shown in fig.\ref{fig:MHV-tree} (b), that is a bubble like structure after attaching two complex scalar together.
 
And finally, the fourth and fifth terms together gives
\bea
\tilde{I}_{\textrm{ll}, T_4+T_5}
%
&=&\frac{\mu^2\,[12]^4 }{[12] [23][34][41] \, [(L-p_1)^2-\mu^2]^2}
-\frac{2 \,\mu^2\,[12]^4\,[42]}{[12] [23][34][41]^2 \langle 1|L|2] \, [(L-p_1)^2-\mu^2]}\nonumber\\
&+&\frac{\mu^2\,[12]^4 \,[42]^2}{[12] [23][34][41]^3 \langle 1|L|2]^2}
\eea

\subsection{Tree-loop factorizations}\label{MHV-TL}
As mentioned in Sec.\ref{MHV-main}, BCFW deformations $|\hat 1]=|1]-z|2],\,\, |\hat 2\rangle =|2\rangle +z|1\rangle$ are applied. 

The factorizations w.r.t. the poles in tree-type propagators contain following contributions to the one-loop integrand 
\begin{itemize}
\item 
First of all, if there is a bubble-like loop structure sitting at an external gluon propagator, this part would give a bubble-like integrand but vanish after the loop-momentum integration when we compute the one-loop amplitude.  
Thus this part can be neglected in the one-loop integrand expression.

\item The contribution from $I(\hat 1^+, \hat P^\mp, 4^-) {i\over s_{23}} A(-\hat P^\pm, \hat 2^+, 3^-)$ and $A(\hat 1^+, \hat P^\mp, 4^-) {i\over s_{14}} I(-\hat P^\pm, \hat 2^+, 3^-)$ is shown in fig.\ref{fig:MHV-TL-2}.

\begin{figure}[!htb]
\centering
\includegraphics[scale=0.9]{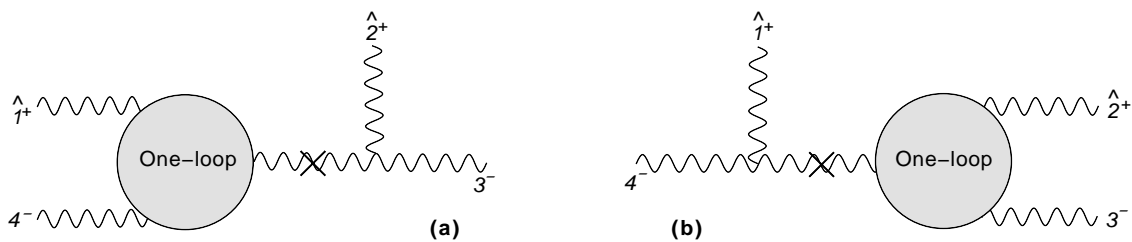}
\caption{MHV tree-loop factorization}
\label{fig:MHV-TL-2}
\end{figure}

Since those two contributions are equivalent to each other, thus we only need to consider one of them, i.e., the one of fig.\ref{fig:MHV-TL-2}(a).

From $s_{\hat 2 3}=s_{23}+z \langle 13\rangle [32]=0$, we obtain $z^*=-\langle 23\rangle /\langle 13\rangle=[41] /[42]$ and  
\bea
&&\,\,|\hat P\rangle = |3\rangle \quad  \rm{and }\quad  |\hat P]= |3]+ {\langle 12\rangle \over \langle 13\rangle} |2]\\
&\rm{equavalently}& \,\, |\hat P\rangle = -\left(|4\rangle+{[12]\over [42]} |1\rangle\right)  \quad  \rm{and }\quad  |\hat P]= |4]
\eea
Then $A(-\hat P^-, \hat 2^+, 3^-)=0$, and the possible non-zero contribution comes from $A(-\hat P^+, \hat 2^+, 3^-)$.

\begin{figure}[!htb]
\centering
\includegraphics[scale=0.9]{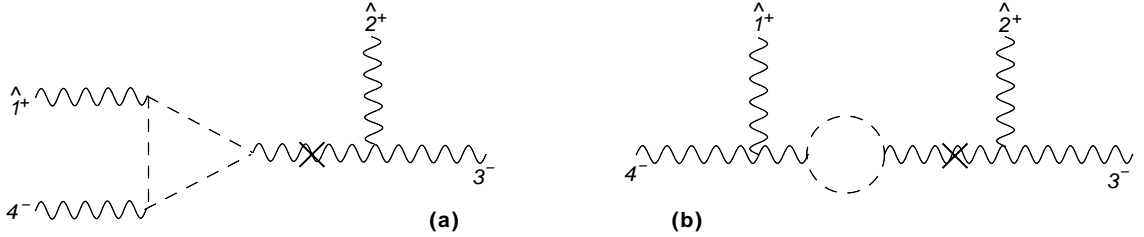}
\caption{MHV tree-loop factorization}
\label{fig:MHV-TL-3}
\end{figure}

On the other hand, $I(\hat 1^+, \hat P^-, 4^-) $ contains both bubble-like and triangle-like integrands, see fig.\ref{fig:MHV-TL-3}. 
Once the on-shellness of $\hat P^-$ holds by $z^*$, the bubble-like integrand fig.\ref{fig:MHV-TL-3}(b), again, can be neglected.
What might contribute to a non-vanishing integrand is from the triangle-like structure in fig.\ref{fig:MHV-TL-3}(a).
 
Take expression of $I_3(1^+, 2^+, 3^-)$ in \eqref{integrand-3pt-oneminus}, flip its helicity and substitute corresponding external particles $\{\hat P^-, 4^-, \hat 1^+\}$ 
\bea
I_3(\hat P^-, 4^-, \hat 1^+)&=&-\frac{1}{ [ 4 \hat 1]} 
\frac{ \langle \hat P |L |\hat 1]^3}{\left[L^2-\mu^2\right]\,\left[(L+\hat p_1)^2-\mu^2\right]\,[(L-p_4)^2-\mu^2]^2}\nonumber\\
&&-\frac{\mu^2\,\langle \hat P \, 1\rangle^3}{\langle \hat P \,4\rangle ^2}\frac{\langle\hat P |L |\hat 1]}{\left[(L+\hat p_1)^2-\mu^2\right]\,[(L-p_4)^2-\mu^2]^3}
\eea
Since $\hat p_1^\mu= p_1^\mu-{[41]\over 2 [42]}\langle 1|\gamma^\mu|2]={[12]\over 2 [42]}\langle 1|\gamma^\mu|4]$, then the above integrand results in a scale-less loop-momentum integration.
\end{itemize}

From the above discussion, we conclude the tree-level factorized channels do not contribute in the 1-loop MHV amplitude.

\subsection{Loop factorizations}\label{MHV-Loop}
In this part, we try to present the computation details in BCFW loop factorizations in a clear and clean way.
As mentioned in Sec.\ref{MHV-main}, the loop-like factorization is constructed from 
\bea
I_{\textrm{MHV}}(z)={i\over (L+\widehat p_1)^2-\mu^2} \tilde{I}_{\textrm{ll,\,MHV}}(L\rightarrow L+p_1,z).
\eea
where $\tilde{I}_{\textrm{ll}, \textrm{MHV}}=\tilde{I}_{\textrm{ll}, T_1}+\tilde{I}_{\textrm{ll}, T_2}+\tilde{I}_{\textrm{ll}, T_4+T_5}$. 
It's better to compute $I_{\textrm{MHV}, T_1}(z)$, $I_{\textrm{MHV}, T_2}(z)$ and $I_{\textrm{MHV}, T_4+T_5}(z)$ respectively, then we might be aware where the non-vanishing integrand up to loop-momentum integration comes from.
\bea\label{Iz_MHV1}
I_{\textrm{MHV}, T_1}\big |_{z=z*}&=&{-i\over (L+\widehat p_1)^2-\mu^2}\frac{2\,\langle 3|L+1|2]^2\,\left\{\,(L^2-\mu^2)\, (\langle 3|L|\hat 1]-\langle 3|4|\hat 1])-\mu^2\langle 3|4|\hat 1]\right\}^2}
{\langle 1|L|2]^2 \,\langle \hat 2\,3\rangle^2\,[4\hat 1]^2\,[(L+p_1+p_2)^2-\mu^2][(L-p_4)^2-\mu^2][L^2-\mu^2]}\nonumber\\
\eea 

With this $z^*$ value which satisfies $(L+\widehat p_1)^2-\mu^2=0$, those following expressions 
\bea
&&\langle 3|L | \hat 1]=\frac{-\mu^2\,\langle 3|1 | 2]- [L^2-\mu^2]\,\langle 3|L+1 | 2]}{\langle 1|L|2]},\quad \langle \hat 2 3\rangle= \frac{\langle 13\rangle}{\langle 1|L|2]} [(L+p_1+p_\gamma)^2-\mu^2],\nonumber\\
&&[ 4\hat 1]\, =- \frac{[42]}{\langle 1|L|2]} [(L+p_1+p_\gamma)^2-\mu^2], \,\, {\textrm{with}} \,\,\, p^\mu_\gamma= \frac{\langle 23\rangle}{2\langle 13\rangle} \langle 1|\gamma^\mu|2]=- \frac{[41]}{2[42]} \langle 1|\gamma^\mu|2]\nonumber
\eea
are substituted into \eqref{Iz_MHV1}.

Most terms will vanish in the loop-momentum integration, due to their scale-less denominators.
A special form required to be carefully dealt with is of the form \eqref{I_scaleless1}.
The integration-non-vanishing terms are
\bea\label{integrand_MHV_T1}
I_{\textrm{MHV}, T_1}\big |_{z=z*}&=&-\frac{2\,i\,\mu^4\,\langle 3|L+1|2]\,\langle 34\rangle^2}{\langle1 2\rangle\, \langle1 3\rangle\,[(L+p_1+p_\gamma)^2-\mu^2]^2\,[(L+p_1)^2-\mu^2][(L-p_4)^2-\mu^2][L^2-\mu^2]}\nonumber\\
&&+\frac{2\,i\,\mu^4\,\langle 34\rangle^2}{ \langle1 2\rangle^2\,[(L+p_1+p_\gamma)^2-\mu^2]\,[(L+p_1)^2-\mu^2]\,[(L-p_4)^2-\mu^2][L^2-\mu^2]}\nonumber\\
&&-\frac{2\,i\,\mu^4\,\langle 34\rangle^2}{ \langle1 2\rangle^2\,[(L+p_1)^2-\mu^2]\,[(L+p_1+p_2)^2-\mu^2][(L-p_4)^2-\mu^2][L^2-\mu^2]}.\nonumber\\
\eea

The other two parts $I_{\textrm{MHV}, T_2}(z)$ and $I_{\textrm{MHV}, T_4+T_5}(z)$ turn out to be 
\bea
&&I_{\textrm{MHV}, T_2}\big|_{z=z*}
=-{i\over (L+\widehat p_1)^2-\mu^2}\bigg\{\frac{2\,\langle 34\rangle^3 \,\langle 3|L+1| 2] \,  \langle 3|L+\hat 1|4]}{\langle 14\rangle\, \langle \hat 23\rangle^3\, [42]\, \langle 1|L|2]\,[(L+p_1+p_2)^2-\mu^2]}\nonumber\\
&&\qquad\qquad\qquad+\frac{\mu^2 \langle 34\rangle^3 \,\langle 21\rangle\,\langle 3|L+1| 2]^2}{\langle 14\rangle\, \langle \hat 23\rangle^3 \,\langle 1|L|2]^2\,[(L+p_1+p_2)^2-\mu^2]^2}\bigg\}\\
~\nonumber\\
&&I_{\textrm{MHV}, T_4+T_5}\big|_{z=z*}=-{i\over (L+\widehat p_1)^2-\mu^2}\bigg\{\frac{\mu^2\,[12]^4 }{[12] [23][34][4\hat 1] \, [L^2-\mu^2]^2}
-\frac{2 \,\mu^2\,[12]^4\,[42]}{[12] [23][34][4\hat 1]^2 \langle 1|L|2] \, [L^2-\mu^2]}\nonumber\\
&&\qquad\qquad\qquad+\frac{\mu^2\,[12]^4 \,[42]^2}{[12] [23][34][4\hat1]^3 \langle 1|L|2]^2}\bigg\},
\eea
which can be verified that they will vanish after loop-momentum integration.

\subsection{Loop-momentum integration}
When we compute the terms in the integrands from loop-like recursion relations, we met a few terms of the form 
\bea\label{I_scaleless1}
I=\frac{\langle 3|L+1|2]^x\,\langle 4|L|4]^y}{[(L+p_1+p_\gamma)^2-\mu^2]^z\,[(L+p_1)^2-\mu^2]\,[(L+p_1+p_2)^2-\mu^2]\,[(L-p_4)^2-\mu^2]}
\eea
with $x>y\geq0$ and $z>0$, which will vanish after loop-momentum integration. 

To demonstrate this nice property to simplify the integrand expression, shift $L\rightarrow L-p_1$ and apply Feynman parameterization on it
\bea
I&=&\frac{\langle 3|L|2]^x\,\langle 4|L-p_1|4]^y}{[L^2-\mu^2]\,[(L+p_\gamma)^2-\mu^2]^z\,[(L+p_2)^2-\mu^2]\,[(L+p_2+p_3)^2-\mu^2]}\nonumber\\
&=&{\Gamma(z+3)\over \Gamma(z)} \int \prod_{i=1}^4 d a_i {\delta(1-\sum_{i=1}^4 a_i) a_2^{z-1} \,{\langle 3|L|2]^x\,\langle 4|L-p_1|4]^y}\over [(L+a_2 p_\gamma +(a_3+a_4) p_2 + a_4 p_3)^2-\mu^2+a_1 a_4 s_{23} ]^{z+3}}.   
\eea
If change the integrated variable from $L$ to $L^\prime=L+a_2 p_\gamma +(a_3+a_4) p_2 + a_4 p_3 $, the integrals of loop momentum after Wick-rotation becomes 
\bea\label{vanish-form-integral}
\mathcal I&=&{\Gamma(z+3)\over \Gamma(z)} \int \prod_{i=1}^4 d a_i \delta(1-\sum_{i=1}^4 a_i) a_2^{z-1} {i\over (-1)^{z+3}}\int {d^D l\over (2\pi)^D} {\langle 3|L^\prime|2]^x\,\langle 4|L^\prime-P^\prime|4]^y\over [l^2-a_1 a_4 s_{23} ]^{z+3}}\nonumber\\
\eea
with $P^\prime=(1-a_4) p_1 +a_2 p_\gamma +a_3 p_2$. 
In the $D=2k-2\epsilon$ dimensional loop momentum integration $ $ with $l^\nu= L^\nu+\mu^\nu$ ($L$ in $2k$-dim and $\mu$ in $-2\epsilon$-dim), integrals with an odd power of the loop momentum in the numerator vanish, while an even power of the loop momentum can be related by Lorentz invariance to scalar integrals together with a combination of $g^{\mu\nu}$, e.g.,
\bea
\int {d^Dl\over (2\pi)^D} l^\mu l^\nu f(l^2) &=&{g^{\mu\nu}\over D} \int {d^Dl\over (2\pi)^D}  l^2 f(l^2)\nonumber\\
\int {d^Dl\over (2\pi)^D} l^\mu l^\nu l^\rho l^\sigma f(l^2) &=&{g^{\mu\nu}g^{\rho \sigma}+g^{\mu\sigma}g^{\rho \nu}+g^{\mu\rho}g^{\mu \sigma}\over D(D+1)} \int {d^Dl\over (2\pi)^D}  (l^2)^2 f(l^2)\nonumber\\
\eea 
and so on for higher loop momentum powers in the numerator.
In our case, since the power of $\langle 3|L^\prime|2]$ $x$ is larger than the power of $\langle 4|L^\prime-P^\prime|4]$ $y$, then the combinations of external spinors of the form $\langle 3|\gamma^\mu|2]\langle 3|\gamma_\mu|2]$ always exist, which results in zero finally.
Due to this fact, the integrals of forms as eq. (\ref{vanish-form-integral}) will always vanish if $x>y\geq0$.

\bibliographystyle{JHEP}

\bibliography{biblio}

\end{document}